¥
\documentclass{eas}
\usepackage{graphicx,natbib}
\usepackage{graphicx,color}

\begin{document}

\title{Multiple stellar populations and their evolution in globular clusters:  \\ A nucleosynthesis perspective} 
\runningtitle{Multiple stellar populations in globular clusters}

\author{Corinne Charbonnel}\address{Department of Astronomy, University of Geneva, Chemin des Maillettes 51, 1290 Versoix, Switzerland \email{Corinne.Charbonnel@unige.ch}}\secondaddress{Institut de Recherche en Astrophysique et Plan\'etologie, CNRS UMR 5277, Universit\'e de Toulouse, 14, Av.E.Belin, 31400 Toulouse, France}
\begin{abstract}
This paper presents a review of the characteristics of the multiple stellar populations observed in globular clusters, and of their possible origin. The current theoretical issues and the many open questions are discussed. 
\end{abstract}
\maketitle

\section{Importance of globular clusters for cosmology and for galactic physics}
\label{GC:importance}

Globular clusters (GC) are magnificent astronomical objects that provide insight to a broad variety of astronomical and cosmological questions.
These compact systems that contain between several hundreds of thousands and a million stars are among the oldest structures in the Universe. 
They are unique relics, witnesses, and clocks of the formation, assembly, dynamics, and evolution of galaxies and of their substructures, from the early to the present-day Universe. 
They play a key role in hierarchical cosmology, and potentially also in the reionization of the Universe, and 
they provide fundamental benchmarks for stellar evolution theory. 
However, their formation, evolution, and survival in a galactic and cosmological context are far from being understood. 

Determining the age of the oldest GCs through isochrone fitting of their observed color-magnitude diagram (CMD) was crucial for cosmology until the beginning of the 21$^{st}$ century, as it provided the best lower limit to the age of the Universe \citep[e.g.][and references therein]{Carney2001}. 
Meanwhile, a robust determination of the age of the Universe was made possible thanks to precise observations of the cosmic microwave background (CMB) by the {\it Wilkinson Microwave Anisotropy Probe} and {\it Planck} satellites.
The age of the universe is now estimated to be  $\sim$ 13.8 according to flat $\Lambda$ CMD model that accurately reproduces the observed temperature of the power spectrum of the CMB \citep{Adeetal_Planck15}. This agrees with the current estimate of the age of the oldest Galactic GCs that were among the first baryonic structures to form in galaxies about 13 Gyr ago \citep[e.g.][]{VandenBerg13}.

Determining the absolute and relative ages of GCs is still very important today, for various reasons. 
This helps constraining the models of the formation of these systems in the more general context of the formation and early evolution of galaxies \citep[e.g.][]{FallRees85,Harris91,Forbesetal1997,vandenBergh2011}. 
Different scenarios make indeed different predictions for the age of the oldest GCs, as well as for the dispersion in age at a given metallicity (or rather [Fe/H]) and/or galactocentric distance. 
There is a wealth of literature on these questions, and we quote here only a few illustrative examples.  
\citet{BrodieStrader} argue that the most metal-poor GCs, which inhabit the haloes of spiral galaxies, were created in low-mass dark matter haloes at redshift higher than 10 (see also \citealt{BrommClarke02}), and that metal-rich GCs formed during later mergers of gas-rich structures that build up the parent galaxies. 
On the other hand, cosmological N-body simulations and semi-analytic models of galaxy and GC formation accounting for the important processes like merging, star formation and feedback, lead \citet{Bekkietal08} to the conclusion that $\sim 90 \%$ of GCs in haloes formed in low-mass galaxies at redshift higher than 3; the mean formation redshifts they deduce for metal-poor and metal-rich GCs are respectively 5.7 and  4.3, which corresponds to modest age difference of $\sim 0.4$~Gyr (i.e., 12.7 and 12.3~Gyr respectively). \citet{Elmegreenetal12} propose that metal-poor GCs could have formed naturally as starbursts in dwarf-star-forming Lyman-$\alpha$ emitting galaxies at intermediate to high redshifts; 
considering the observed mass-metallicity relation in galaxies, they argue that the formation of the metal-poor GCs ``(...) is not the exclusive result of an earlier birth time compared to metal-rich disk and bulge GCs, but rather the result of a lower mass host".
More recently, \citet{Kruijssen2015} discuss the properties of present-day GCs and show that they can be reproduced by assuming that they are the natural outcome of regular high-redshift star formation. Also, first steps are being made in order to understand GC formation in
cosmological simulations \citep{KravtsovGnedin05,Boleyetal09,Griffenetal10}. 

Despite its importance, the determination of the absolute ages of Milky Way GCs remains problematic. 
Although the very powerful method of model isochrone fitting to observed CMDs has been used for decades to age-date these large sample of stars  \citep[e.g.][]{VandenbergBell85}, 
it is still limited by the uncertainties inherent to stellar evolutionary models, but also to the error on apparent distance modulus, on reddening, and on the relations between colors and effective temperature, bolometric correction scale, etc  \citep[e.g.][and references therein]{CassisiEES2013,CasagrandeVandenBerg14,LebretonetalEES2014}. 
Recent studies using the Main Sequence Turnoff (MSTO) fitting method for deep HST CMDs agree on an age for the oldest metal-poor GCs (MPGC) of 12.5~Gyr \citep{VandenBerg13} to 12.8~Gyr \citep{MarinFranchetal09}, which is consistent with the cosmological age inferred from the measurements of the expansion rate of the Universe. 
However, there is no consensus between the most recent studies on the age of the metal-rich GCs. On one hand, \citet{MarinFranchetal09} find that 
the oldest metal-rich GCs (MRGC) are roughly co-eval with the MPGCs,  while \citet{VandenBerg13} find that they are 
systematically younger with a mean age of 11~Gyr  (this excludes the much younger MRGCs that follow a different age-metallicity relation and are associated with accretion events, \citealt{ForbesBridge2010}).  

At the time of massive acquisition of very precise photometric, astrometric, spectroscopic, and asteroseismic data with large ground-based and space surveys which goal is to  decipher the assembly history of our Galaxy by mapping the structure, chemical composition, dynamics, and age distribution of its stellar populations\footnote{In the Gaia era we can quote spectroscopic surveys such as RAVE, SEGUE, APOGEE, LAMOST, GALAH and GaiaESO, photometric surveys such as OGLE, CRTS, PanSTARRs, SkyMapper and the future LSST, and the asteroseismic missions CoRoT, Kepler, K2, and future PLATO}, 
the computation of state-of-the-art stellar models as well as 
reliable transformations between the observational and theoretical planes remain the major building blocks and the main challenges in the field. 
GCs today still provide some of the best constraints to these necessary developments.

\section{Importance of globular clusters for stellar physics}
\label{GCsforstellarphysics}

\subsection*{From the classical paradigm ...}

Because of their age, only low-mass stars are still alive (thermonuclearly speaking) in GCs today; the typical turnoff mass is of the order of 0.8 - 0.9~M$_{\odot}$, depending on the metallicity and the age of the cluster. In the CMD they populate the main sequence (MS; central hydrogen burning phase), the subgiant and the red giant branches (RGB; shell-hydrogen burning phase around the helium degenerate core), the horizontal branch (HB; central helium burning phase), the asymptotic giant branch (AGB; double shell burning phase around the carbon-oxygen degenerate core), and the white dwarf cooling sequence. 

It was long taken for granted that every GC is the perfect archetype of a single stellar population.
In other words, the stellar hosts of individual GCs were believed to be coeval (i.e., born in a single star formation burst) with different initial masses distributed according to an initial mass function (IMF), 
and to have all formed with the same chemical composition (in terms of both helium and metal abundances).  
Indeed, while the range in [Fe/H] covered by Galactic GCs is very large, with values between $\sim$ -2.3 and 0 (e.g.  \citealt[e.g.][]{Harris10,Carrettaetal09c,Diasetal2016}), nearly all GCs appear to be fairly homogeneous in heavy elements \citep[i.e., Fe-peak, neutron-capture, and alpha-elements, see
  e.g.][]{James04,Sneden05,Carrettaetal09c}, with the notable exception of the most massive ones like 
  $\omega$ Cen (\citealt{Butleretal1978}; note that it is not considered as a {\it{bona fide}} GC but rather as the remnant of a dwarf galaxy), M22, M54, or NGC~3201 
  \citep[e.g.][ and references therein]{DaCosta_M22_09,JohnsonPilachowski10,Siegel07,Carrettaetal10M54,Simmereretal2013}.
According to this ``classical paradigm", GCs  were thought to have undergone no internal chemical evolution (their metals were not produced in situ, but were inherited by the proto-GC from progressive metal-enrichment during the halo chemical evolution; e.g. \citealt{HarrisPudritz1994,James04}),
nor any dramatic dynamical events that could have led e.g. to a drastic modification of their mass along their life (see H.Baumgardt, this volume, for a review on GC dynamics).
With this definition, a single stellar population is defined by a single isochrone in the CMD.

Therefore, with thousand hundreds to millions of stars of similar age and metallicity that remained bound together since their formation, individual GCs have long been recognized as the ideal laboratories for the physics of low-mass stars 
\citep[e.g.][]{Sandage1953,ArpJohnson1955,Sandage1958,SandageWallerstein1960,RoodIben1968,IbenRood1970,Osborn1971PhD}. 
The developments of stellar evolution theory took advantage in particular of the comparison of the observed positions of the stars in clusters' CMDs with the predicted paths of the stellar evolution models in the Hertzsprung-Russell diagram (HR; despite the challenge raised by the transformations between the observational and the theoretical planes). 
This approach has guided the development of many concepts in stellar evolution, among which the explanation of the morphological variety of the horizontal branch, many details of the evolution along the asymptotic giant branch and the white dwarf cooling sequence, or the nature of blue stragglers. 
It has provided some of the best constraints to stellar models in the pre-asteroseismic era and 
significantly helped improving the input macro- and micro-physics physics of the models of low-mass stars. 
In particular, it lead to some of the most significant progresses in the understanding and the description of convection, atomic diffusion, rotation-induced mixing, hydrodynamical instabilities, equation of state, opacities, mass loss, model atmosphere, and nuclear reactions in stellar interiors \citep[e.g.][ and references therein]{VandenBergetal02,LebretonetalEES2014}. 
Last but not least, the large range in [Fe/H] covered by Galactic GCs provided crucial tests on the impact of metallicity on stellar evolution as well as anchors for metallicity scales (e.g. \citealt{KraftIvans03,Carrettaetal09c}).

\subsection*{... to the recent revolution}

One of the major breakthroughs of the past decade in stellar population studies concerns those massive star clusters (MSC) we thought we knew the best, namely GCs.  
Indeed and at odds with the classical paradigm described above, we know now that these old systems actually host multiple (at least two) stellar populations (MSPs). 
Those exhibit very peculiar chemical properties 
and leave unexpected imprints in the clusters color-magnitude diagrams 
that have never been observed yet in any other stellar population, neither in the field, nor in open clusters. 
This discovery has totally revolutionized the field. In \S~\ref{section:multiplepopulationsobsspectroscopy} 
we describe the phenomenology of GC MSPs and use that term for stars that can be distinguished within individual GCs either from their spectra or from their positions in the CMD.

\section{Multiple stellar populations in globular clusters - Chemical properties}
\label{section:multiplepopulationsobsspectroscopy}

The number of spectroscopic studies devoted to the chemical analysis of GCs has considerably increased over the past couple of decades, and it is out of the scope of this paper to provide an exhaustive list of all the publications in the domain. 
Rather, we select representative analysis to illustrate the main observational features and refer to the review by \citet{Gratton12}  for an extensive reference list.

\subsection{Carbon, Nitrogen, Oxygen, Sodium, Magnesium, Aluminium, Silicium, Potassium}
\label{HBproducts}

After the discovery of two N-enhanced stars in M5 and M10 \citep{Osborn1971PhD},  different distributions of CN and CH band strengths\footnote{Since molecular abundances are controlled by the abundances of the minority species, CH and CN trace carbon and nitrogen abundances respectively.}
 were obtained with low-resolution measures among both main sequence and giant stars in several GCs \citep[e.g.][]{Zinn77,Hesser80}. These striking findings prompted the following tremendous spectroscopic efforts devoted to the chemical dissection of GCs. 
Pioneer high-resolution studies were limited to the brighter red giant stars. 
They revealed {\bf{systematic star-to-star variations in the abundances of}} 
\begin{itemize}
\item {\bf{carbon, nitrogen, oxygen}} \citep[e.g.][]{Carbonetal82,Pilachowskietal83,PaltoglouNorris89,Brownetal90,Snedenetal91,Kraftetal92},  
\item {\bf{sodium}} \citep[e.g.][]{Cohen78,Peterson80},   
\item {\bf{aluminium}}, and sometimes {\bf{magnesium}} \citep[e.g.][]{Norrisetal81,BrownWallerstein92,NorrisDaCosta95,Shetrone96}
\end{itemize}

in all the individual GCs under investigation. 
Importantly, the abundances of C and N \citep[e.g.][]{Butleretal1978},  Na and O \citep[e.g.][]{Peterson80,CottrellDaCosta81,Drakeetal92,Kraftetal93}, and Mg and Al \citep[e.g.][]{Ivansetal01,Snedenetal04} appeared to be anticorrelated to different extent depending on the cluster. 

A major step forward resulted from the advent of high-resolution spectroscopy on 8-10 meter class telescopes, which allowed the chemical dissection of turnoff main sequence stars in the brightest Galactic GCs\footnote{Using low-resolution spectra, \citet{Brileyetal96} had already discovered variations in Na correlated with CN in MSTO stars in 47~Tuc.}. 
After the seminal papers by \citet{Gratton01} and \citet{Theveninetal01}, the large star-to-star light-element abundance variations of C, N, O, Na, Mg, and Al, coupled with {\bf{characteristic (anti-)correlations (e.g., C-N, Na-O, Mg-Al)}} have been convincingly detected down to the subgiant branch and the MSTO \citep[e.g.][]{RamirezCohen02,RamirezCohen03,Carrettaetal04,Lindetal09,D'Orazietal10}. 
This brought compelling evidence that GCs host multiple stellar populations {\bf{born with unique initial chemistry}}, as discussed in details in  \S~\ref{section:nucleosyntheticorigin}.

\begin{figure}
\centering
\includegraphics[width=12cm]{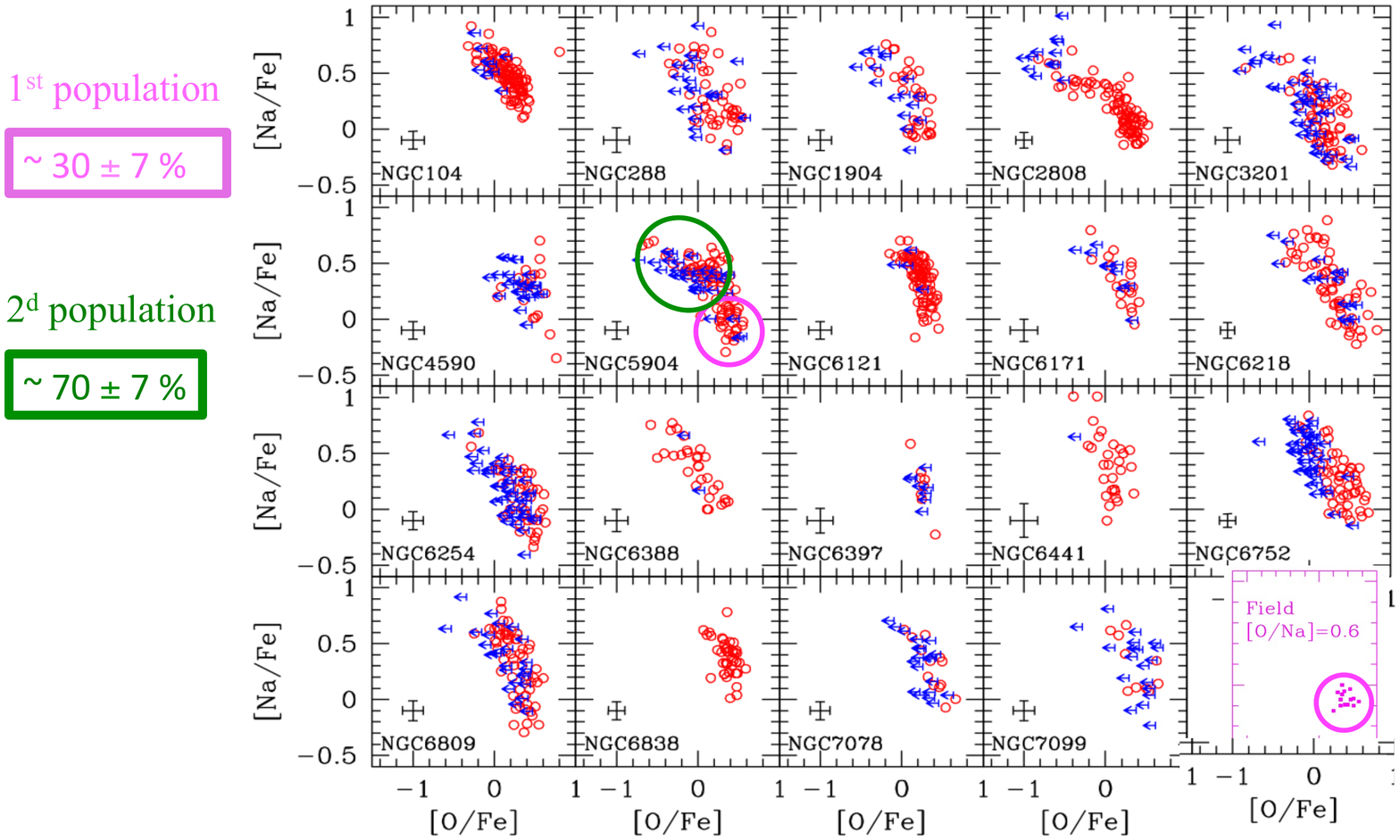}
\caption{O-Na anticorrelation observed in nineteen GCs by \citet[][GIRAFFE and UVES spectra]{Carrettaetal09b}.  
Upper limits in O abundances are shown as blue arrows, detections are indicated as red circles.
The magenta circle in the  NGC~5904 panel indicates the area in the O-Na plane where stars have abundances similar to those of field halo stars, while the green ellipse surrounds the Na-enriched, O-depleted stars. This corresponds respectively to the so-called first and second stellar populations. 
We also show (bottom right panel) the O and Na abundances in field metal-poor stars with -2$\leq$[Fe/H]$\leq$ -1 by \citet{Gratton00}. Figure adapted from \citet{Carrettaetal09b}}
\label{Fig:ONaanticorrelationCarretta09VII}
\end{figure}

Figure~\ref{Fig:ONaanticorrelationCarretta09VII} shows the O-Na anticorrelation observed in nineteen Galactic GCs with FLAMES/UVES on VLT by Carretta and collaborators. The comparison with data for field stars indicates the abundances with which the proto-GC clouds must have formed. The corresponding ``normal" GC stars (with respect to halo field star composition) are called {\bf{first population}} GC stars (hereafter {\bf{1P}}, or first generation), while the Na-enriched, O-depleted GC stars are called  {\bf{second population}} ({\bf{2P}}, or second generation) stars.

\begin{figure}[h]
\centering
\includegraphics[width=0.6\textwidth]{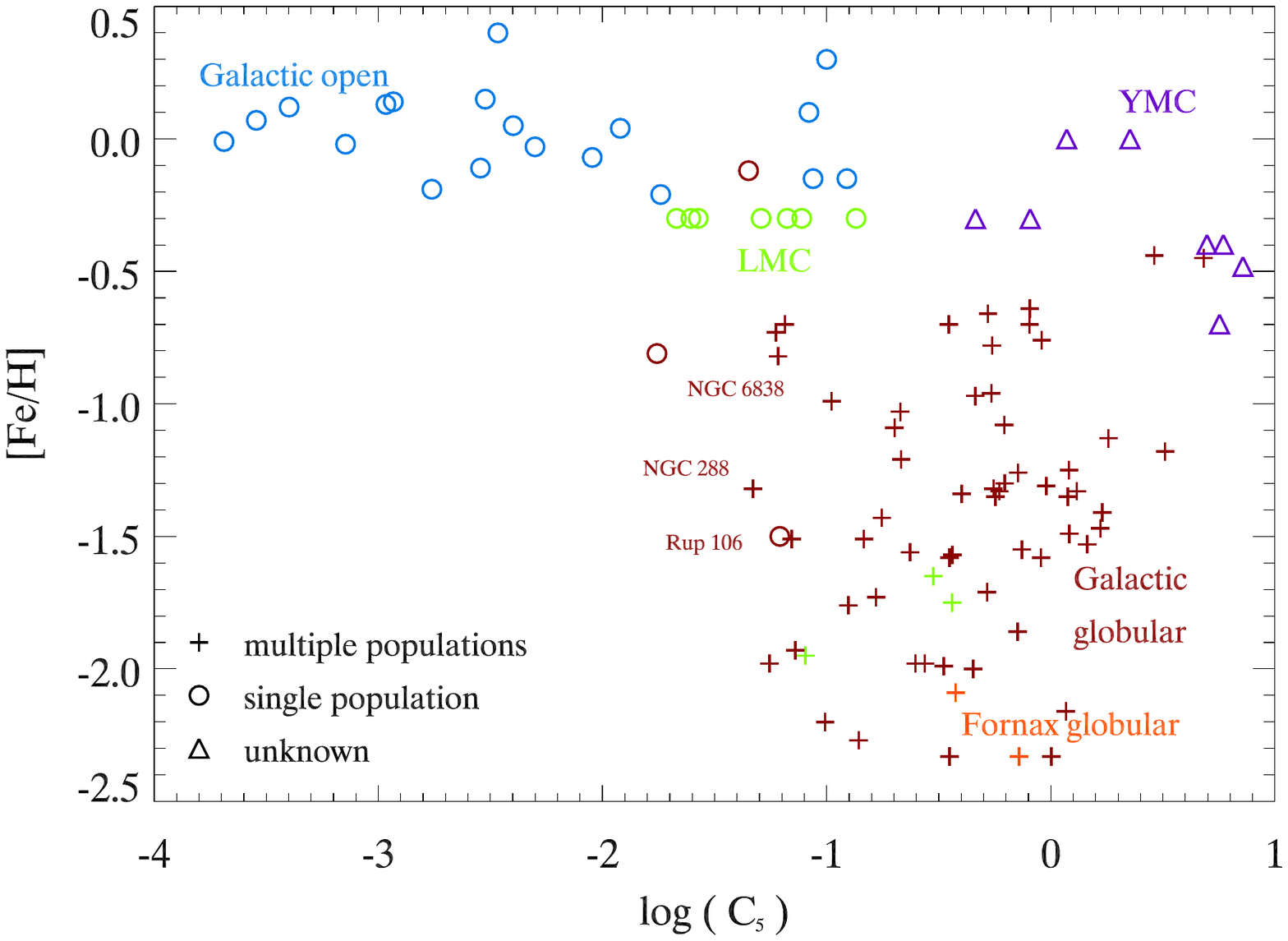} 
\caption{Star clusters of various types and origins (GCs in brown and orange, young massive star clusters in purple, and open clusters in green) in the [Fe/H] -- compactness plane. Different symbols indicate the absence or presence of the O-Na anticorrelation (circles and plus respectively; triangles for clusters where the
presence of the anticorrelation is unknown). Multiple populations occur at high C$_5$ where gas expulsion becomes increasingly difficult and eventually impossible due to large binding energies. See \citet{Krauseetal16} for more details and for sample references. Figure from \citet{CharbonnelKrause_iaus317}}
\label{fig:CCKrause_IAUS317}
\end{figure}

All the Milky Way GCs surveyed to date show an O-Na anticorrelation. 
The shape of this feature -- i.e., its extension expressed by the interquartile ratio (\citealt{Carretta06}) and its slope -- varies from cluster to cluster and seems to depend on the absolute magnitude of the GCs, which reflects their total mass \citep{Carrettaetal09b}. 
A present-day mass larger than a few 10$^4$~M$_{\odot}$ is presently estimated to be the lower limit for a GC to exhibit the O-Na anticorrelation \citep{Carrettaetal10}. 
This is corroborated by the fact that this pattern has never been found yet in the less massive, less compact open clusters \citep{deSilvaetal09,Pancinoetal10,Bragagliaetal12,Bragagliaetal14}, with the possible exception of the old, very massive, and very-metal rich galactic open cluster NGC~6791  \citep{Geisleretal12}. Star clusters with and without the O-Na anticorrelation actually separate clearly in a compactness-metallicity diagram \citep{Krauseetal16}, as shown in Fig.~\ref{fig:CCKrause_IAUS317} where the compactness index C$_5$=(M$_*$/10$^5$M$_{\odot}$)/(r$_h$/pc), with  M$_*$ the stellar mass and r$_h$ the half-mass radius of each individual cluster.

\begin{figure}[h]
\centering
\includegraphics[angle=-90,width=0.8\textwidth]{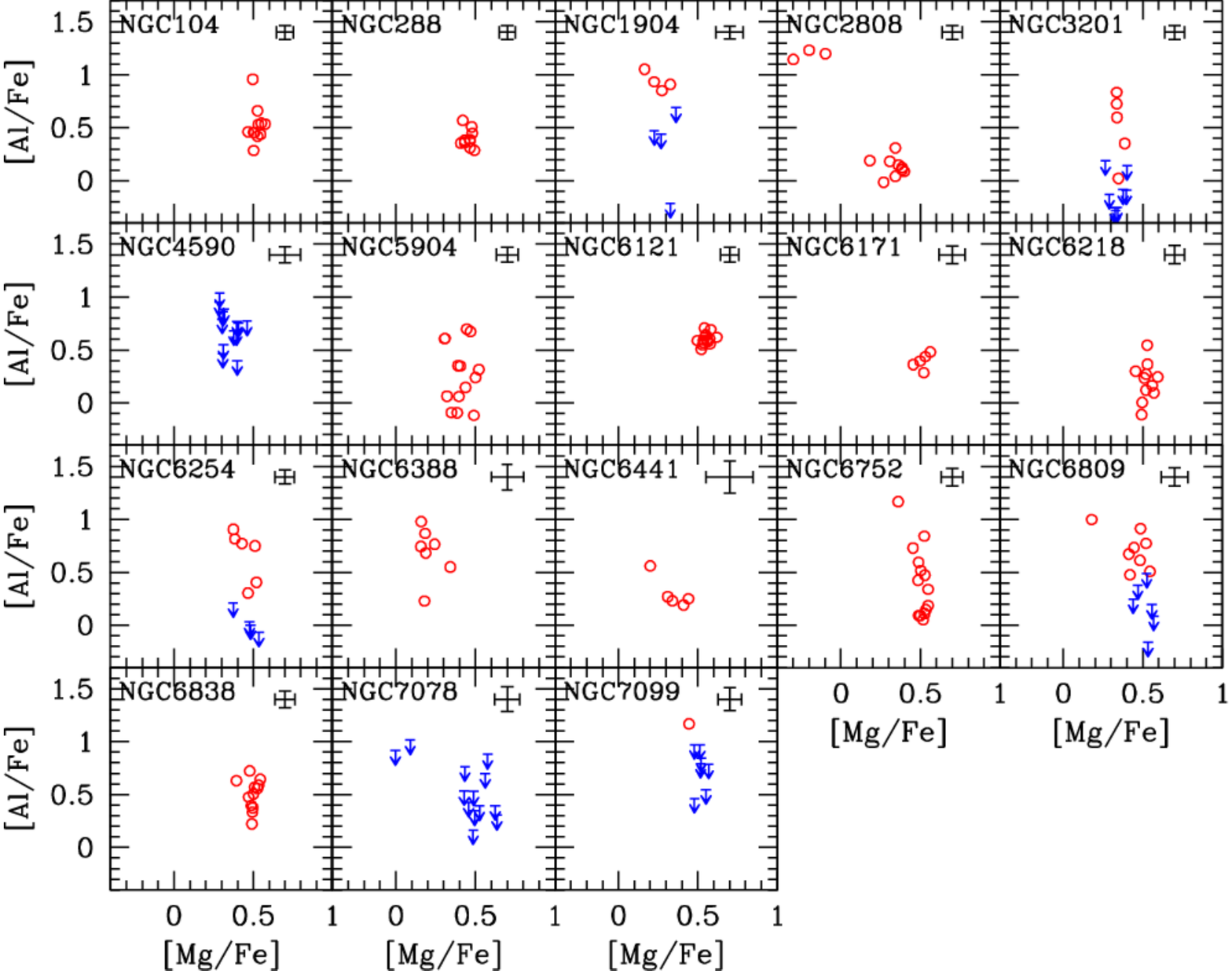} \\
\includegraphics[angle=-90,width=0.8\textwidth]{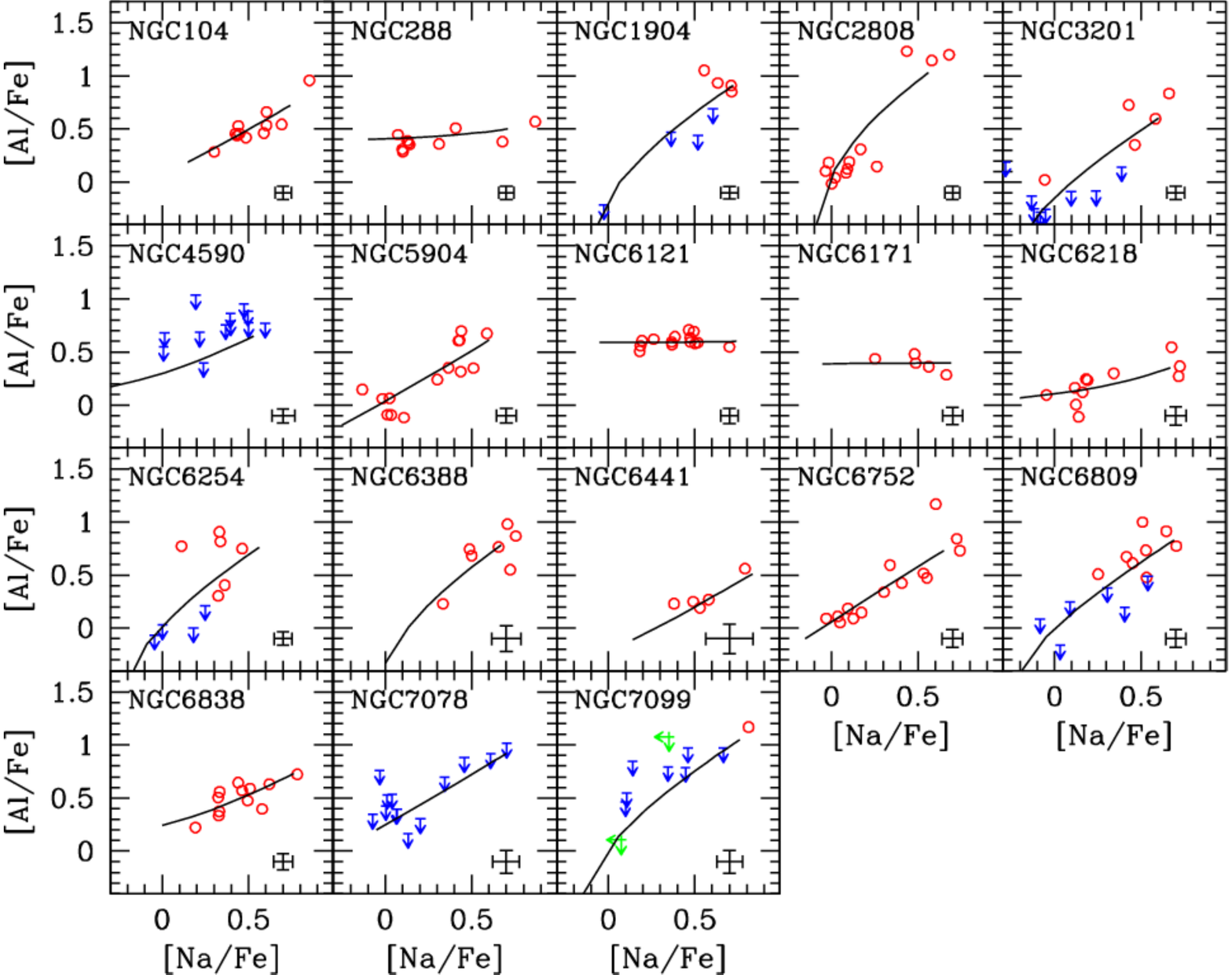}
\caption{[Al/Fe] ratios as a function of [Mg/Fe] and [Na/Fe] ratios from UVES spectra observed in 18 of the 19 GCs shown in Fig.~\ref{Fig:ONaanticorrelationCarretta09VII}. Detections are shown as circles, upper limits as arrows. Star-to-star error bars are indicated in each panel. Figures from \citet{carrettaetal09a}}
\label{Fig:MgAl_Carrettaetal09VIIIfig6}
\end{figure}

\begin{figure}
\centering
\includegraphics[width=0.5\textwidth]{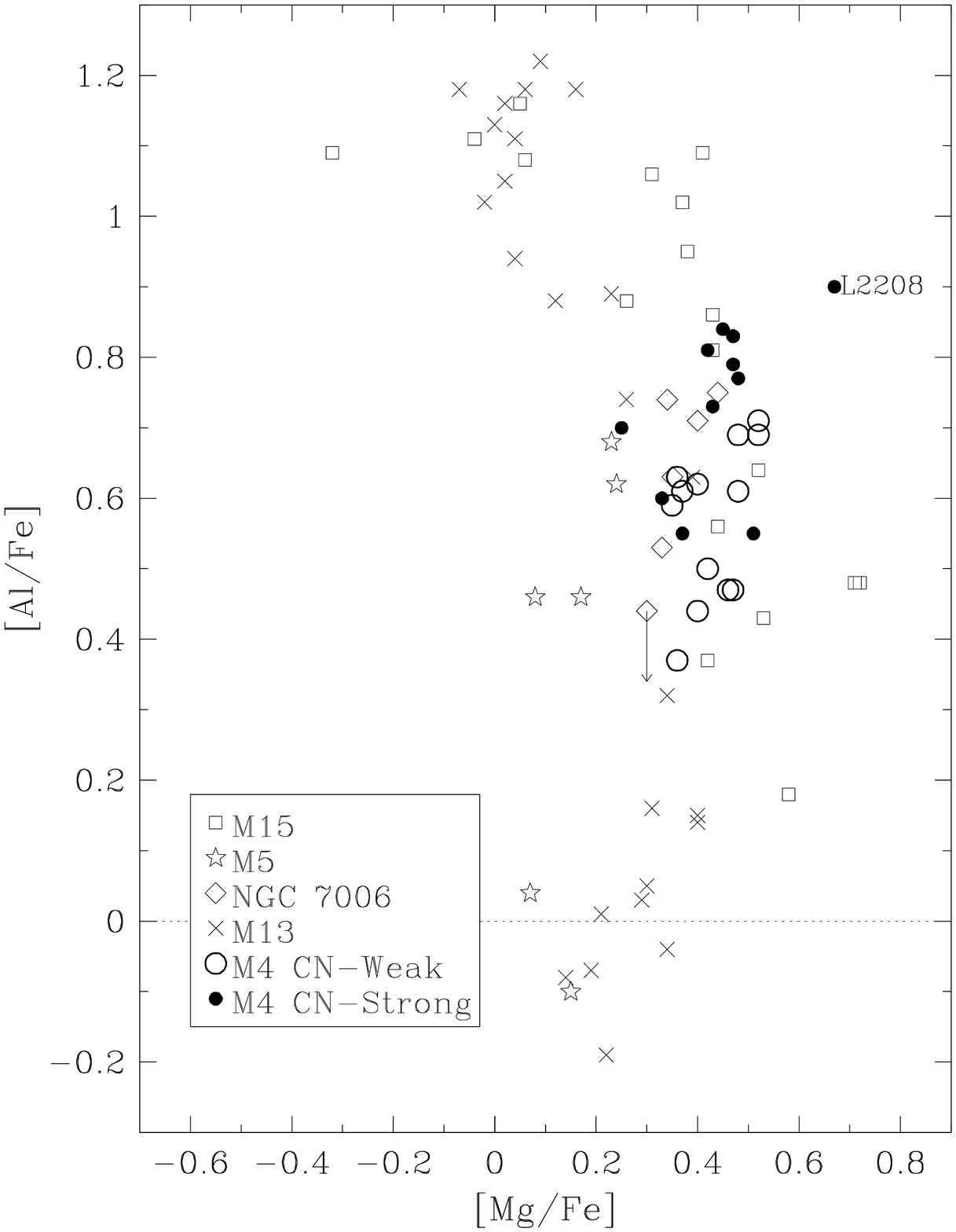}
\caption{Aluminium versus magnesium abundances from high-resolution analysis of stars in five GCs. 
In some clusters like M4, there is no correlation between Al and Mg, whereas other GCs show either correlated or anti-correlated behaviours. Figure from \citet{Ivansetal99}}
\label{Fig:MgAl_Ivansetal99}
\end{figure}

Aluminium is the heaviest chemical species displaying large star-to-star variations in Galactic GCs (Fig.~\ref{Fig:MgAl_Carrettaetal09VIIIfig6}; e.g.
\citealt{Norrisetal81,Ivansetal01,Gratton01,Johnsonetal05,Johnsonetal08,Carrettaetal13Alu,Corderoetal15}). 
An Al-Na correlation is found in metal-poor GCs (M13, M5, M80, NGC~1851, NGC~2808, NGC~6752; \citealt{Carretta12Alu}) but not in the metal-rich ones (47~Tuc and M71; \citealt{Corderoetal14,Corderoetal15}). Some GCs, but not all, show a weak anticorrelation between Mg and Al (Fig.~\ref{Fig:MgAl_Carrettaetal09VIIIfig6} and \ref{Fig:MgAl_Ivansetal99}).
Finally, an extremely unusual Mg-K anticorrelation was discovered in NGC~2808 \citep{Mucciarellietal15} and in the old and metal-poor cluster GC NGC~2419  \citep{Cohenetal11,CohenKirby12,Mucciarellietal12} which is the most massive GC orbiting in the outermost suburbs of the halo of the Milky Way. 

The abundance variations of C, N, O, Na, Mg, and Al observed in GCs are clearly depicting the results of  {\bf{hydrogen-burning through the CNO-cycle and the NeNa- and MgAl-chains}}, as will be discussed in details in  \S~\ref{section:nucleosyntheticorigin}. 
Importantly, the {\bf{C+N+O abundance sum }} is found to be constant (within the measurement uncertainties) among stars of individual GCs (e.g. \citealt{Carrettaetal05}), except in the case of NGC~1851 (\citealt{Yongetal15}, but see \citealt{Villanovaetal10}) and M22 (\citealt{Marinoetal12}) which are both very special GCs showing substantial star-to-star abundance scatter in iron-peak and neutron-capture elements (\S~\ref{Iron-rs-abundances}).

\subsection{Helium} 
\label{obshelium}

There are several indications for helium abundance variations among GC cluster stars. Evidence comes both from 1) high-precision photometry: 
multiple main sequences or subgiant branches (e.g. \citealt{Piotto07,Milone08}),  extended  horizontal branches (e.g. \citealt{Miloneetal14}), or changes in the mean Na and O abundances at the RGB bump that can be interpreted as the result of the impact of helium on the bump luminosity (e.g. \citealt{Salaris06}), and 2) from direct measurements of helium abundances in RGB and HB stars (e.g. \citealt{Dupreeetal11,Pasquinietal11,DupreeAvrett13,Marino14}).
As of today, the maximum helium spread on helium determined by isochrone fitting is  0.13~dex in NGC~2808 that exhibits a triple MS \citep{Miloneetal12}.

\subsection{Lithium} 
\label{LiBeB}

The photospheric lithium abundance of GC stars begins to be documented in a systematic way. 
However, the interpretation of the observations is not straightforward and it has to be done with great caution.
Lithium is indeed a fragile element that burns at relatively low temperatures in stars ($\sim 2.5 \times 10^6$~K). 
In the case of the Sun, the initial abundance at the time of the formation of our star could be determined precisely thanks to measurements in meteorites (e.g. \citealt{Loddersetal09} for a review), and it is $\sim 180$ times higher than the photospheric Li abundance in the present-day Sun \citep{Asplundetal09}. 
Sophisticated stellar models can reproduce the solar lithium depletion as well as the lithium behavior in open cluster solar-type stars (e.g. \citealt{Palaciosetal03,CT05}). 
However, these models are not yet fully predictive because of the degeneracy of the problem due to not-well established interplay between several complex mechanisms that can decrease the internal and surface Li content of low-mass stars at different phases of their evolution: rotation-induced mixing, atomic diffusion, mass loss, internal gravity waves, magnetic fields, thermohaline mixing, as well as dilution during the first dredge-up episode on the RGB. Moreover, there is both theoretical and empirical evidence that the respective efficiency of these mechanisms depends on stellar mass, metallicity, and rotation (e.g. \citealt{DPC00},  \citealt{PCD00},  \citealt{CDP00}, and more recent reviews in the proceedings of IAU Symposium 268, \citealt{Charbonnel08_IAUS268}; also e.g.  \citealt{Ramirezetal12Li,Tognellietal12,Castroetal16,Carlosetal16,DelgadoMenaetal15}). 
Therefore, while these processes should in principle be taken into account self-consistently in order to derive the initial abundance with which the stars were born, it is still impossible today to disentangle all the effects  (e.g. \citealt{Zahn13,Basuetal15,Palacios15}) and to trace back the initial Li content of individual stars. 

\begin{figure}[h]
\centering
\includegraphics[angle=-90,width=\textwidth]{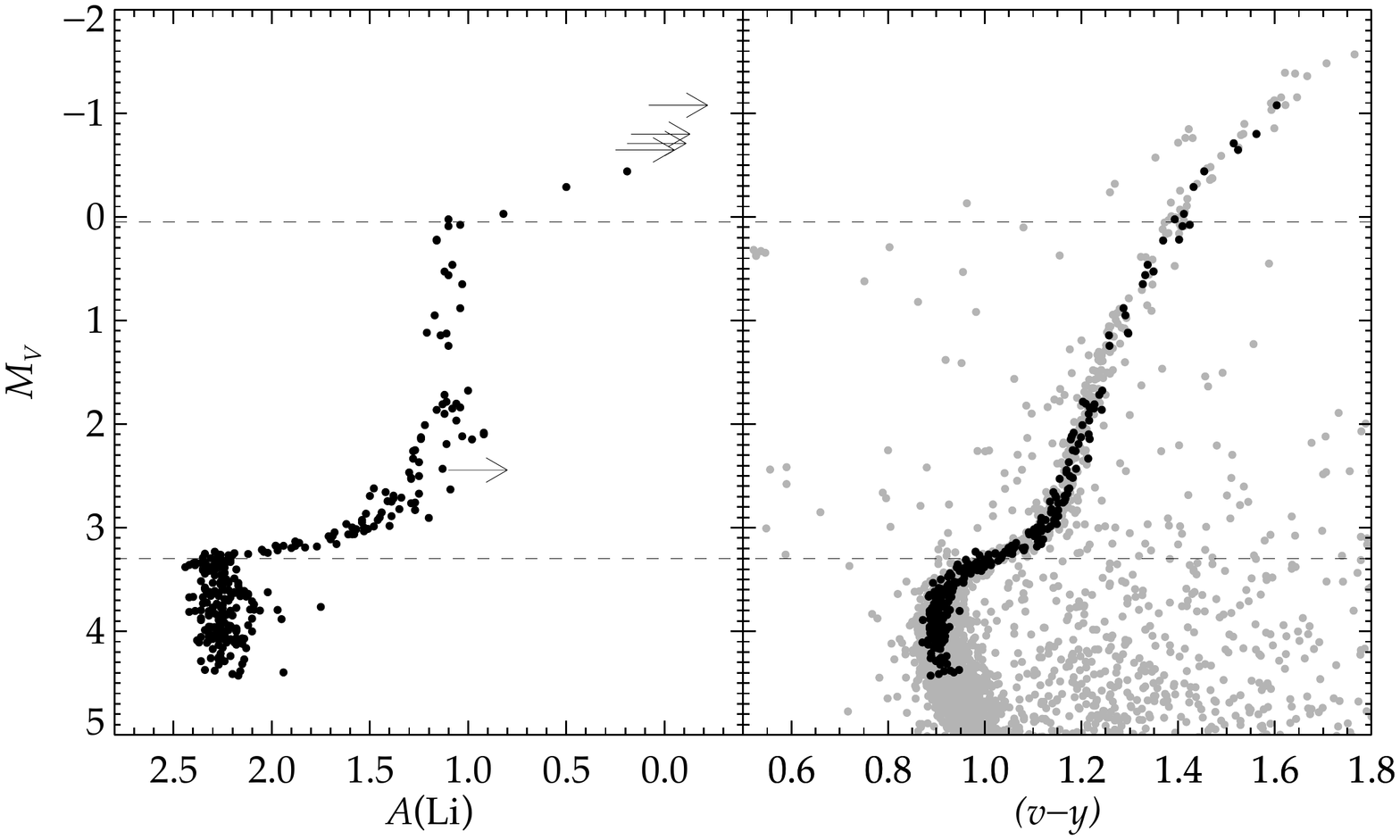}
\caption{Non-LTE lithium abundance (left) for spectroscopic targets in the CMD of NGC~6397 (right). The dashed lines indicate the positions of the MS turnoff (Mv$\sim$3.3) and of the RGB bump (Mv$\sim$0.05) where the first dredge-up event and thermohaline mixing respectively lead to lithium depletion. Figure from \citet{Lindetal09}}
\label{Fig:Li_NGC6397_Lind2009}
\end{figure}

An upper limit for the initial Li content of metal-poor halo stars and of proto-GC gas comes from the predictions of Standard Big Bang Nucleosynthesis constrained by cosmic microwave background experiments (e.g. \citealt{Coc14}). The value of this so-called primordial or cosmological Li abundance is A(Li) between 2.659 and 2.728 (uncertainty due to the uncertainties on nuclear reaction rates; A(Li)=12+log[N(Li)/N(H)]) using the baryonic density $\Omega_b h^2 $= 0.02207 ({\it Planck}, \citealt{Planck13}).
This is about two to three times higher than the maximum Li abundance derived for  MS turnoff stars in GCs (NGC~6752, \citealt{Pasquini05}; 47~Tuc, \citealt{Bonifacioetal07}; M~30, \citealt{Gruyters16}; NGC~6397, \citealt{Bonifacioetal02,Kornetal07,Lindetal09}) as well as for halo MS stars (e.g. \citealt{Gratton00,CharbonnelPrimas05,Sbordoneetal10}). 
The representative case of NGC~6757 is shown in Fig.~\ref{Fig:Li_NGC6397_Lind2009}: The dwarf, turn-off, and early subgiant stars (with Mv higher than $\sim$ 3.3)
are located along an abundance ``plateau" at the level of the so-called Spite plateau for field stars. This plateau is interrupted in the middle of the subgiant branch by the Li dilution due to the occurrence of the first dredge-up event. A second steep abundance drop is seen at the luminosity of the red giant branch bump (Mv$\sim$0.05) when thermohaline mixing occurs (\citealt{CharbonnelZahn07a}). 
Therefore, the evolution of Li along the stellar life of GC hosts is very similar to that observed in Galactic halo stars, and they do show similar signatures of intrinsic stellar depletion processes (see e.g. \citealt{Gratton00,CharbonnelPrimas05}). 

\begin{figure}[h]
\centering
\includegraphics[angle=-90,width=\textwidth]{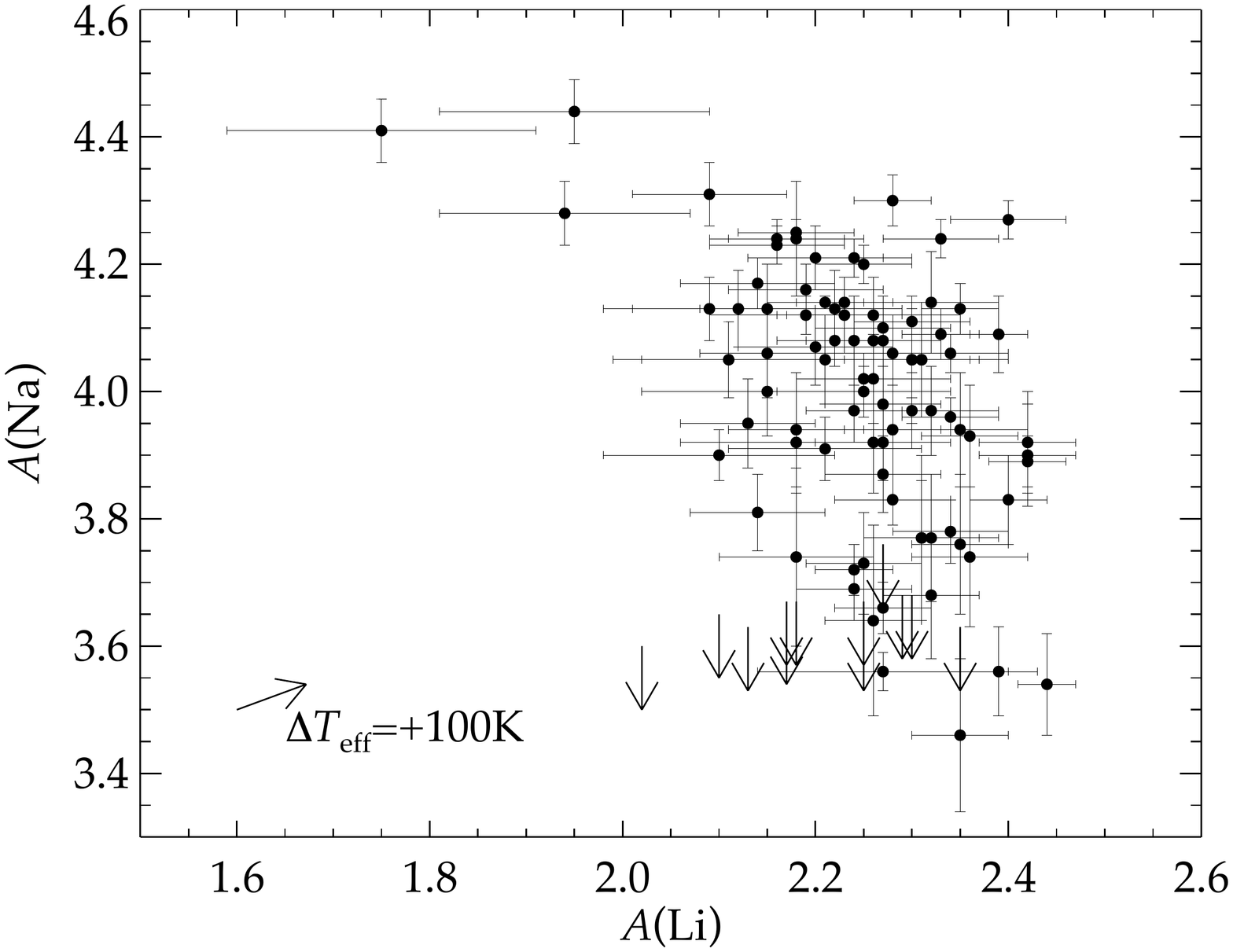}
\caption{Non-LTE lithium and sodium abundance in dwarf and subgiant stars in NGC~6397 (arrows indicate Na upper limits). Figure from \citet{Lindetal09}}
\label{Fig:LivsNa_NGC6397_Lind2009}
\end{figure}

Additionally, the spread in Na and Li abundances is very large when one considers GC MS and turnoff stars that have not yet undergone Li dilution due to dredge-up (about one order-of-magnitude in the case of NGC~6397; Fig.~\ref{Fig:Li_NGC6397_Lind2009}). The most Li-poor GC MS turnoff and subgiant stars also show the highest sodium abundances (Fig.~\ref{Fig:LivsNa_NGC6397_Lind2009}), which is indicative of an anticorrelation (which significance is under debate) between the abundances of Li and Na. Such abundance pattern is a strong constraint to the internal chemical evolution of GC, as will be discussed in \S~\ref{section:nucleosyntheticorigin}.

\subsection{Iron-peak and neutron-capture elements}
\label{Iron-rs-abundances}

\begin{figure}[h]
\centering
\includegraphics[width=0.7\textwidth]{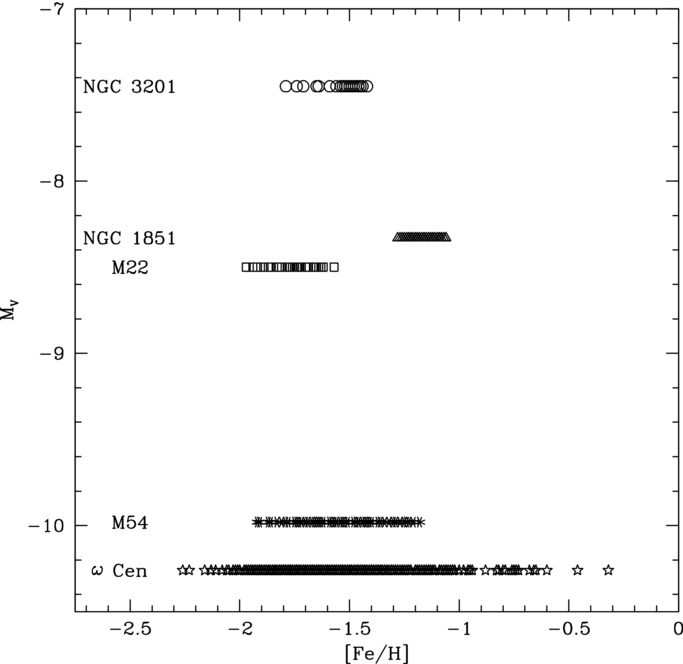}
\caption{GC integrated magnitude (M$_V$) for metal-poor Galactic GCs with internal iron abundance variations confirmed with high-resolution spectroscopy. Figure from \citet{Simmereretal2013}}
\label{Fig:ironspread}
\end{figure}

The large majority of Galactic GCs are fairly homogeneous in iron-peak and neutron-capture elements (e.g. \citealt{Gratton04,Carrettaetal09c}). 
However, star-to-star [Fe/H]  spreads have  been found in the most massive GCs, both in the Milky Way and in dwarf spheroidal galaxies. 
The most extreme and puzzling case is $\omega$~Cen, which [Fe/H] covers more than 1.5~dex (with discrete components; e.g. \citealt{NorrisDaCosta95IV,Villanovaetal14}) and that also shows abundance variations of neutron-capture elements (e.g. \citealt{Marinoetal11}). 
$\omega$~Cen, which is the  most luminous and most massive Milky Way GC ($\sim 3 \times 10^6$~M$_{\odot}$), may however be an outlier among MW GC population, as it has long been suspected to be the stripped remnant of an accreted dwarf galaxy (e.g. \citealt{Leeetal99,Meylan03}). 
The other GCs that show evidence of supernovae (SNe) enrichment in addition to the spreads in
light elements (however much lower than in $\omega$~Cen) are M54 that is associated with the nucleus of Sagittarius dwarf spheroidal galaxy (e.g. \citealt{LaydenSarajedini97,Bellazzinietal08,JohnsonPilachowski10,Carrettaetal10M54}), NGC~3201 \citep{GonzalezWallerstein98,Simmereretal2013}, NGC~1851 \citep{Carretta12Alu}, M22 \citep{DaCosta_M22_09,Marinoetal12,Lee09OmCenM22}, Terzan~5 \citep{Ferraroetal09}, and NGC~6273 \citep{Johnsonetal16}. Figure~\ref{Fig:ironspread} shows the [Fe/H]  spread in some of these special GCs as a function of their integrated magnitude that is an indicator of their total stellar mass. In some of these GCs ($\omega$~Cen, NGC~1851, and M~22), indications of internal dispersion of neutron-capture elements have also been found (e.g. \citealt{Marinoetal09,Roedereretal11,Yongetal14}).

\section{The nucleosynthetic origin of the abundance anomalies}
\label{section:nucleosyntheticorigin}

\subsection{Hydrogen-burning in stars through the CNO cycle and the NeNa- and MgAl-chains}
\label{subsection:CNONeNaMgAl}

\begin{figure}
\centering
\includegraphics[width=0.45\textwidth]{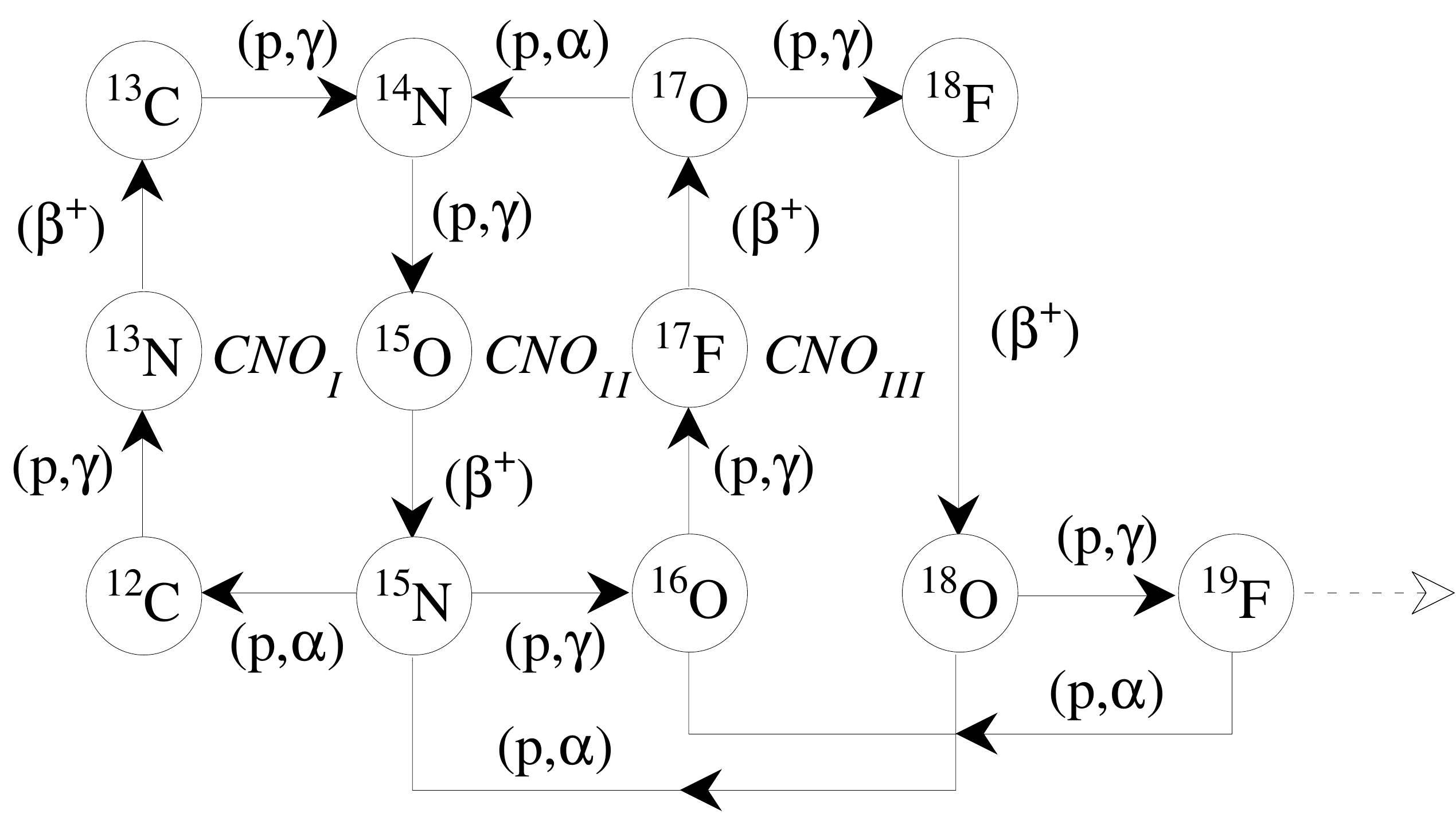}
\includegraphics[width=0.45\textwidth]{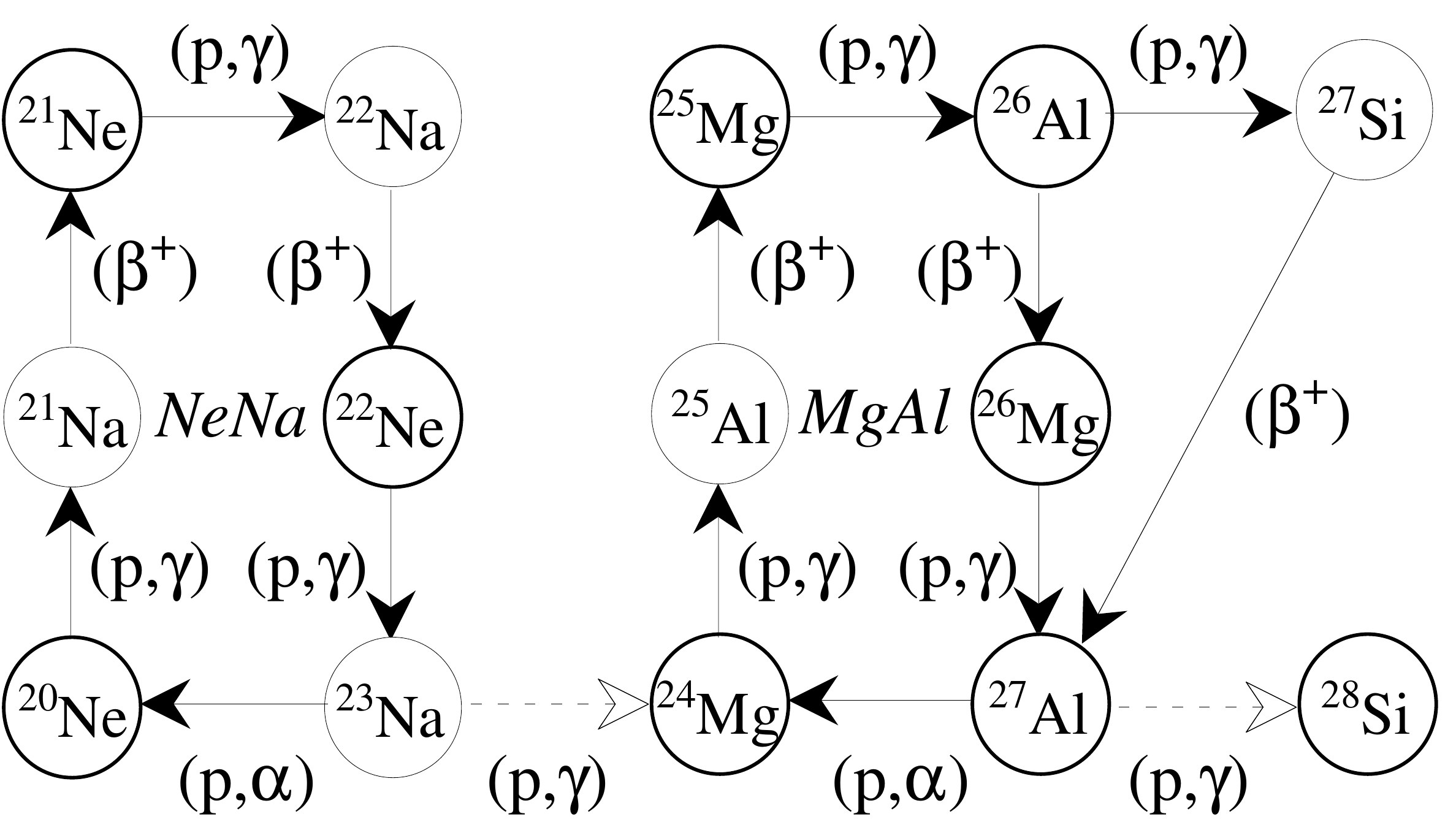}
\caption{Proton-captures through the CNO-cycle and the NeNa- and MgAl-chains. Figures from \citet{Arnould99}}
\label{Fig:CNONeNaMgAl}
\end{figure}

The C-N, O-Na, Mg-Al anti-correlations that we described in \S~\ref{section:multiplepopulationsobsspectroscopy} can be unambiguously understood as the result of H-burning through the CNO-cycle and the NeNa- MgAl-chains \citep{DenisenkovDenisenkova90,Langeretal93,Prantzos06,PCI07}.
In stellar conditions proton-captures on $^{16}$O and $^{22}$Ne lead respectively to the destruction of $^{16}$O and the production of $^{23}$Na at temperatures of the order of 25~MK.
At higher temperatures of $\sim$ 40~MK, $^{20}$Ne also generates $^{23}$Na. 
The equilibrium value of $^{23}$Na is highest at $\sim$ 50~MK, and it decreases at higher temperature.
Importantly, the Mg-Al chain becomes active at higher temperature and the two most fragile isotopes $^{25}$Mg and $^{26}$Mg are the first ones to be destroyed.
$\sim$ 70~MK are required to start depleting $^{24}$Mg. Substantial amounts of $^{27}$Al are produced through this chain.  
Finally, K can be produced through a succession of reactions starting from $^{36}$Ar at temperatures of the order of 180 -- 200~MK \citep{Iliadisetal16}. In such extreme conditions, Na is destroyed; this implies that Na and K must come from different sources (Prantzos, Charbonnel \& Iliadis, in prep).
For more details on H-burning in stellar conditions and on the uncertainties on the nuclear reaction rates we refer e.g. to \citet{Arnould99} and \citet{Xuetal13NACRE}.

\begin{figure}[h]
\centering
\includegraphics[width=0.7\textwidth]{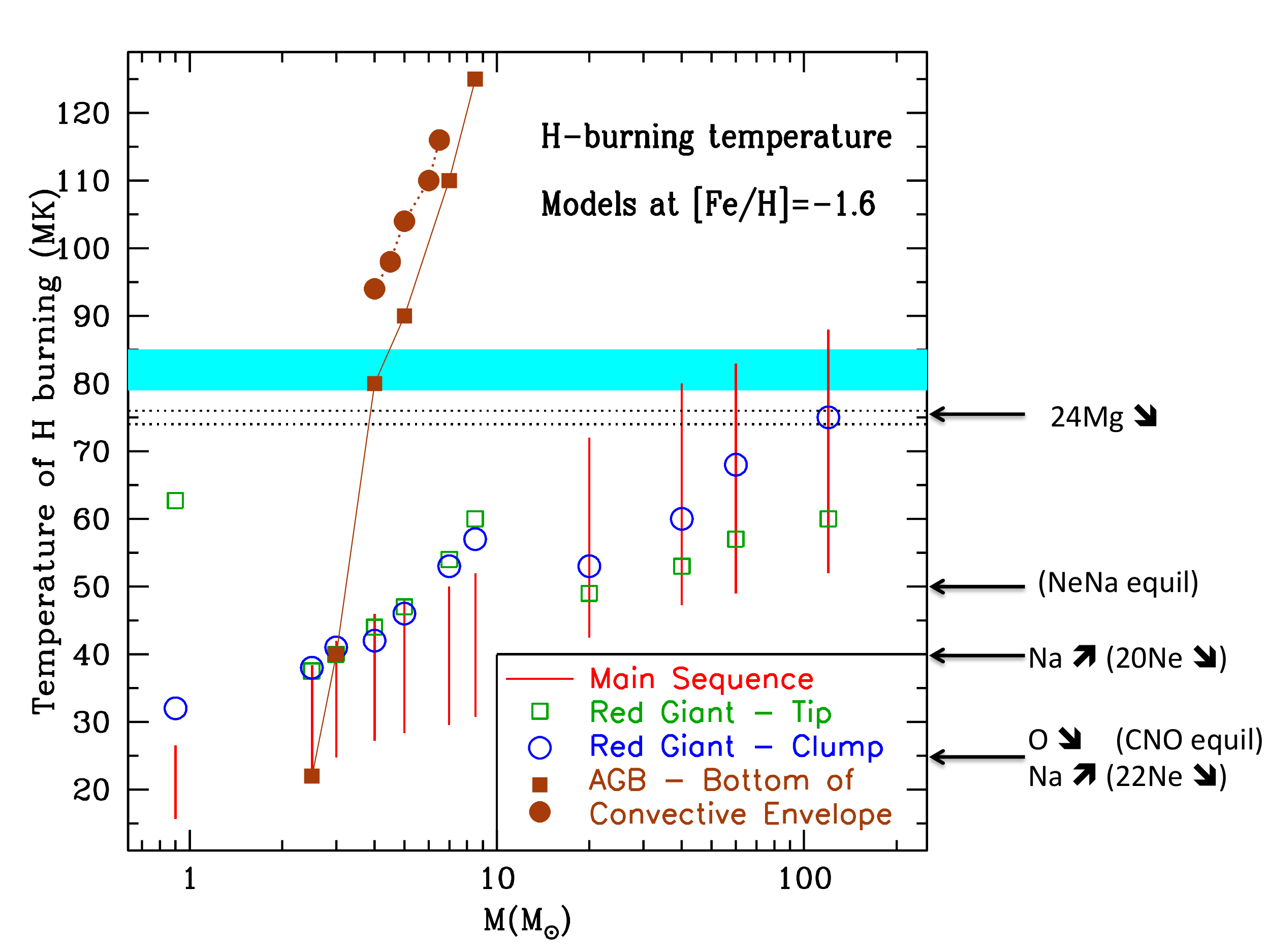}
\caption{Temperature of hydrogen burning in stars as a function of their initial mass and at various phases of their evolution (symbols in insert) for [Fe/H]~=~-1.75. 
Central H-burning temperature are given from the zero age main sequence to the turnoff; for the other evolution phases only the maximum H-burning temperature is given.
On the left of the figure, the arrows indicate the temperature for the efficient activation of proton-capture on different chemical elements and their isotopes.
Figure adapted from \citet{PCI07} 
}
\label{Fig:PrantzosCI06fig6}
\end{figure}

The maximum temperature for H-burning in stars of different masses and at different phases of their evolution is given in Fig.~\ref{Fig:PrantzosCI06fig6} for the case of stars with [Fe/H]~=~-1.75 (for any given mass the corresponding temperature would be higher for lower values of [Fe/H]). 
In the case of core H-burning we show the range covered by the central temperature during the whole main sequence. 
For the other phases we show only the maximum temperature reached in the H-burning regions of interest : in the H-burning shell (HBS) during the red giant phase (i.e., 
at the ``tip" of the RGB) and during the central He-burning phase (i.e, on the ``clump"), and at the bottom of the convective envelope in AGB and super-AGB stars. 

\subsection{Origin of the abundance anomalies: Nature or nurture?}
\label{subsection:Hburningtemperaturel}

Two explanations to the abundance patterns observed in GCs were originally proposed in the literature. 

\subsection*{Nurture -- Stellar evolution effects}
At the time when abundances were determined for GC giant stars only, it was proposed that GC stars could eventually modify their surface composition by in situ processes along the RGB. 
Several theoretical works investigated the possibility of the operation of very deep mixing between the stellar convective envelope and the HBS that surrounds the degenerate He-core of RGB stars
(e.g. \citealt{DenisenkovDenisenkova90,Kraftetal93,Langeretal93,LangerHoffman95,Pilachowskietal96,CavalloNagar00,Weissetal00}). 
When low-mass stars climb the RGB indeed, H-burning temperature increases in the shell  so that the CNO-cycle operates at equilibrium there (Fig.~\ref{Fig:PrantzosCI06fig6}).
At the tip of the RGB, the temperature becomes sufficient in the HBS  
for the NeNa-chain to operate and for p-captures to affect the two heavier Mg isotopes (but not to deplete $^{24}$Mg; e.g. \citealt{Powelletal99}).  

As a matter of fact, it is firmly established from the observational point of view that the surface abundances of lithium, carbon, and nitrogen, as well as the carbon-isotopic ratio, do change well after the first dredge-up in low-mass red giant stars when they reach the luminosity bump on the RGB, both in the field and in globular clusters (see e.g. the Li decrease at M$_V \sim 0.05$ in Fig.~\ref{Fig:Li_NGC6397_Lind2009}; \citealt{Gratton00,Shetrone03,Lindetal09}). 
These observations of C, N, and Li can be well explained by the development of thermohaline mixing (also called double-diffusive instability, or fingering convection) induced by the molecular weight inversion created by the $^3$He($^3$He,2p)$^4$He reaction in the HBS (\citealt{Kippen80,CharbonnelZahn07a,CharbonnelLagarde10}).
Although the efficiency of this mixing process in stars is still under debate (e.g. \citealt{Denissenkov10,DenissenkovMerryfield10,Traxleretal11,Brownetal13,GaraudBrummell15}),  it should not reach deep and hot enough layers to induce O and Na variations in RGB stars. This is in agreement with the fact that no sign of sodium and oxygen abundance variations could be found at that phase in field and open cluster stars \citep{Gratton00,Smiljanicetal16} nor in GCs\footnote{In the peculiar case of M~13 however, there are some indications that the most O-poor stars stand at the tip of the RGB \citep{CohenMelendez05,JohnsonPilachowski12}, which could probe a marginal evolutionary effect.}.  Additionally, low-mass red giant branch stars never reach internal temperatures high enough to activate the Mg-Al chain (see \S~\ref{section:candidatepolluters} and Fig.~\ref{Fig:PrantzosCI06fig6}).

The main evidence against the historical argument relating in situ deep mixing in the course of stellar evolution and the O-Na and Mg-Al anticorrelations was 
the discovery in the early 2000's of the same chemical anomalies down to the MSTO in GCs \citep{Gratton01,Theveninetal01}.  
As can be seen in Fig.~\ref{Fig:PrantzosCI06fig6}, the H-burning temperature in the turnoff stars that we are currently observing in GCs and that have masses of 
the order of 0.8 - 0.9~M$_{\odot}$ is too low for the NeNa- and MgAl-chains to operate. 
Therefore, {\it{the observed abundance variations cannot be created within these low-mass main sequence stars themselves, and the so-called evolution explanation has to be discarded}}. 

\subsection*{Nature -- GC self-enrichment}

Today, we are thus left with the so-called {\it{primordial explanation}} according to which 
GC stars must have inherited most of their chemical peculiarities at birth (e.g. \citealt{Peterson80,BrownWallerstein92}). 
Different {\bf {``self-enrichment" scenarios}} have been developed in the literature, most of them calling for the formation of at least two stellar populations in all GCs during their infancy. 
In this framework, first population (1P) stars are thought to be born with the proto-cluster original composition (i.e., that of contemporary field halo stars, with typical SNeII paterns), while second population (2P) stars formed from original gas polluted to various degrees by hydrogen-burning processed material ejected by more massive, short-lived, 1P GC stars (the ``polluters")\footnote{Importantly, GC abundance anomalies are of similar magnitude in MS and RGB stars that have respectively very narrow and very extended convective envelopes. Therefore, the polluters' ashes can not have been simply accreted onto the surface of 1P stars, but they must have been mixed with pristine gas before 1P stars formed.}.

To identify the  {\it{candidate polluters}}, we must look at the maximum temperature for H-burning in stars of different masses and at different phases of their evolution (Fig.~\ref{Fig:PrantzosCI06fig6} for [Fe/H]~=~-1.75). 
We discuss the different options in \S~\ref{section:candidatepolluters}.

\section{Candidate stellar polluters for GC self-enrichment - The nucleosynthesis point of view}
\label{section:candidatepolluters}

Different models of secondary star formation have been proposed to explain cluster self-enrichment by short-lived ($\leq$ 100 million years), massive ($\geq$ 6~M$_{\odot}$) stars that have long disappeared in GCs. The proposed possible polluters are asymptotic giant branch stars 
(AGB, $\sim$6.5~M$_{\odot}$; e.g. \citealt{Ventura01,Ventura13,Decressin09a,D'ercole12,D'Antonaetal16}), 
fast rotating massive stars (FRMS, $\geq$ 25~M$_{\odot}$; \citealt{Maeder06,Prantzos06,Decressin07a,Decressin07b,Krauseetal13}), and supermassive stars ($\geq 10^4$~M$_{\odot}$; \citealt{DenissenkovHartwick14,Denissenkovetal15}). 
The possible contribution of intermediate-mass binaries was also suggested (10 -- 20~M$_{\odot}$;  \citealt{DeMink09})
as well as that of FRMS paired with AGB stars or with high-mass interactive binaries \citep{Sills10,Bastianetal2013a}.

As of today, none of these scenarios (at least in their current form) is able to reproduce the extreme variety of the observed nucleosynthetic patterns and to reconcile all the spectroscopic and the photometric signatures of GC self-enrichment. 
Here we summarize their pros and cons in terms of nucleosynthesis, while other aspects of the associated global scenarios will be discussed in \S~\ref{section:selfenrichmentscenario}.
We refer to other recent publications for a more general discussion on all the other aspects related to the formation and to the dynamics of massive star clusters in a galactic and cosmological context (e.g. \citealt{Bastianetal15,Charbonnel_iaus316,Renzinietal15,Krauseetal16}).

\subsection{AGB stars - Nucleosynthesis}
\label{subsection:AGBnucleosythesis}

The first stellar sources that have been proposed to be at the origin of GC self-enrichment were   
massive AGB stars ($\sim$ 5 -- 10~M$_{\odot}$).
In their pioneer paper \citet{CottrellDaCosta81} proposed that the Na and Al enrichment observed in 
CN-strong stars of the GCs 47~Tuc and NGC~6752 was produced by neutron-captures on $^{22}$Ne and $^{25}$Mg within the thermal pulse (the neutrons being released by the $^{22}$Ne($\alpha$,n)$^{25}$Mg reaction; \citealt{Iben76}). Subsequent mass loss was then invoked to release these chemicals in the remaining proto-cluster gas from which Na-enriched, CN-strong 2P stars form. 
Then \citet{D'Antonaetal83} suggested that the spreads in C and N observed in GCs could result from pollution of the external layers of already existing low-mass stars thanks to accretion of matter lost by intermediate-mass stars (2.5 to 4.5~M$_{\odot}$; see also \citealt{Ventura01,Ventura02,Thouletal02}). However, the accretion scenario had to be discarded after the discovery of the fact that MS and red giants present the same ranges in abundance anomalies. 

Massive AGB stars present several qualitative advantages that made them initially very attractive for a self-enrichment scenario. 
They possibly experience the so-called hot bottom burning (hereafter HBB) that processes the material included inside their large convective envelopes through CNO, NeNa and MgAl nucleosynthesis. They do not produce $\alpha$- or Fe-peak elements. 
Their low-speed winds may potentially be retained within the cluster.
Finally, their lifetime is relatively short, $\sim$ 50 - 100~Myr depending on the initial stellar mass and metallicity.  
This is however longer than the lifetime of more massive stars that explode as SNeII and which yields are not observed in most GCs. 
Consequently, in the AGB scenario all the pristine material that remained after the formation of 1P stars must be expelled out of the GC together with 1P SNe ejecta before 2P stars can form; as discussed in more details in \S~\ref{subsection:AGBscenario}, this implies that external pristine material was re-accreted by the GC after the SNe phase to form 2P stars.

\begin{figure}[h]
\centering
\includegraphics[width=0.4\textwidth]{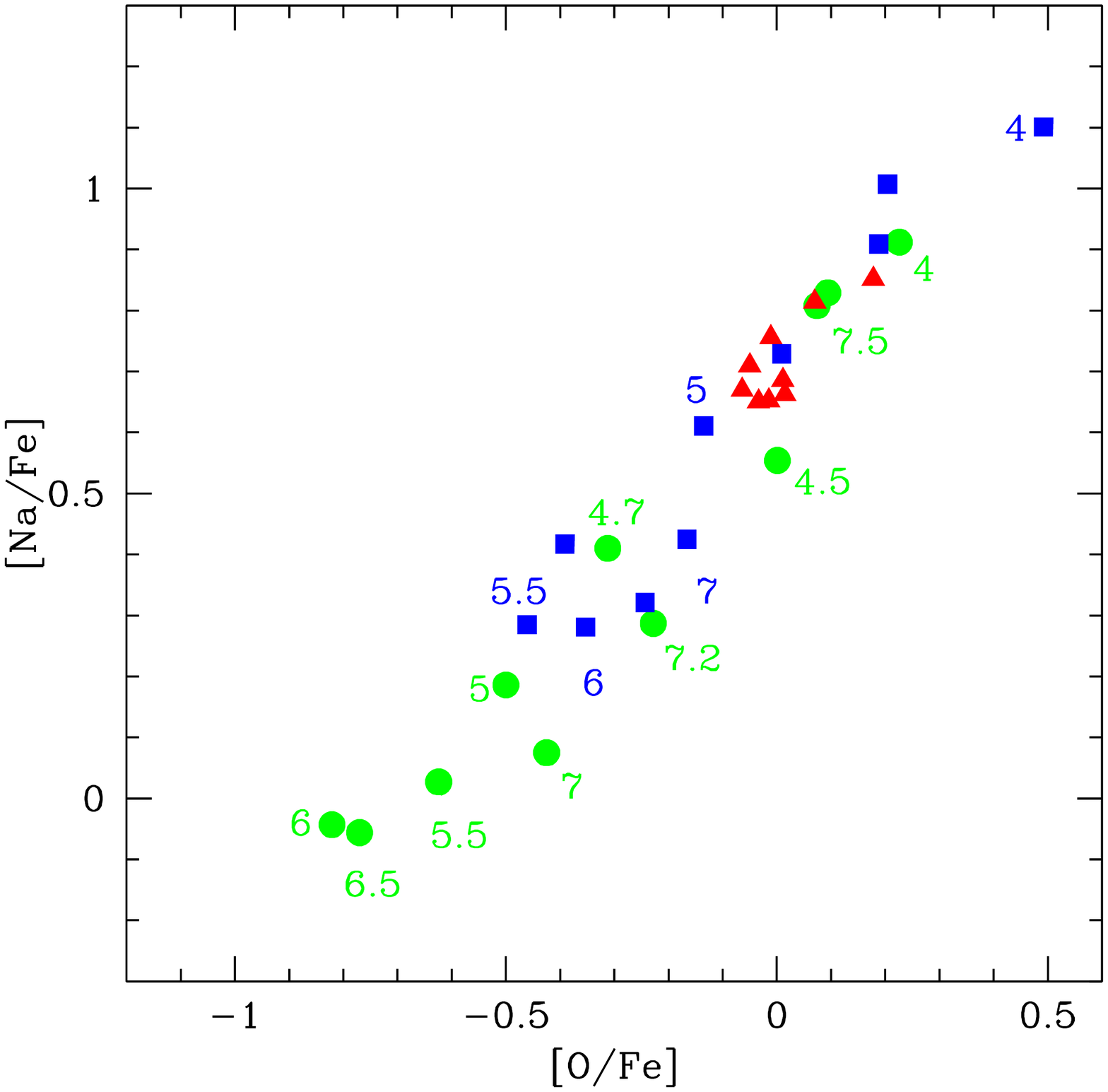}
\includegraphics[width=0.4\textwidth]{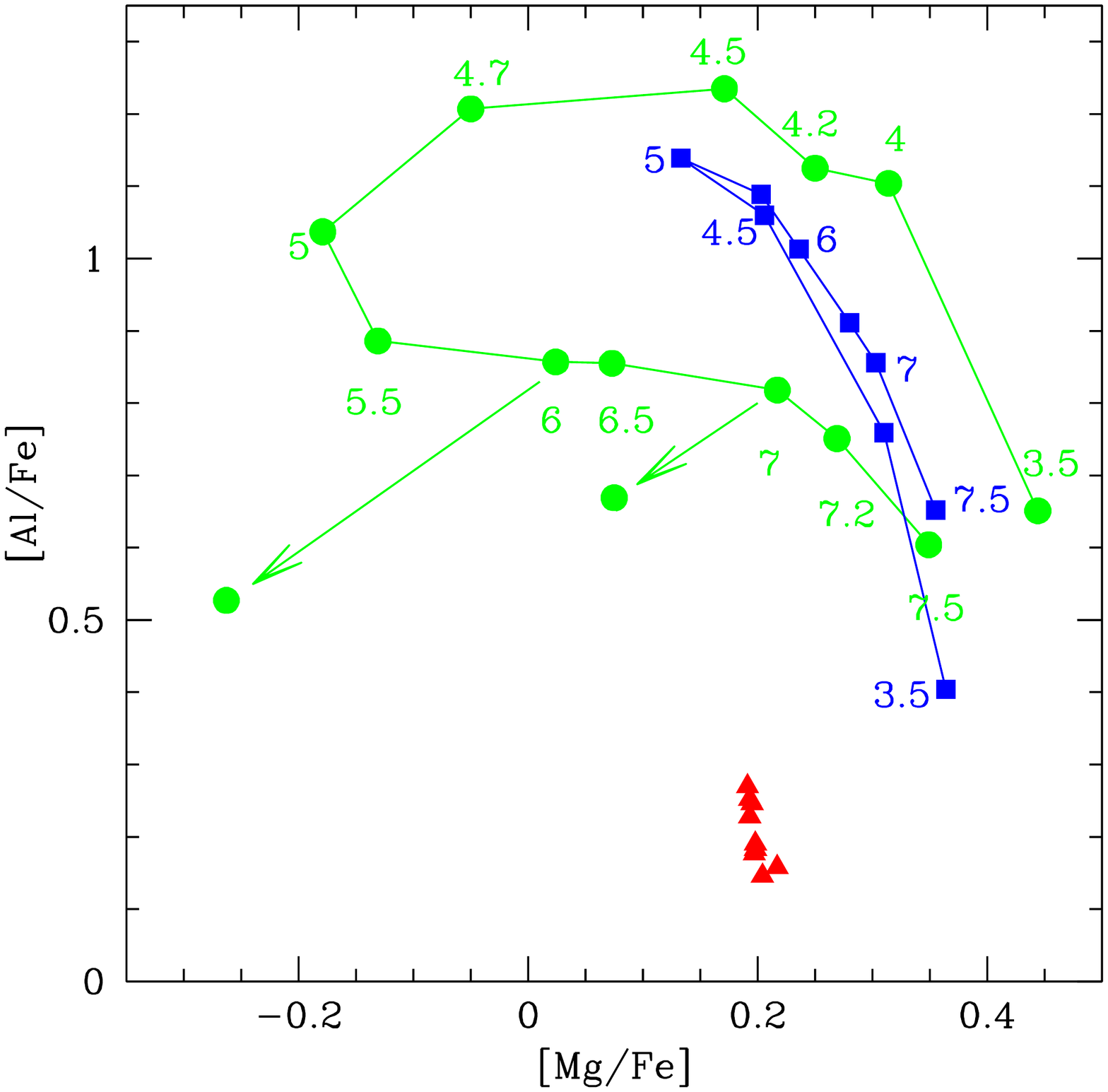}
\caption{Yields in the OÐNa and MgÐAl planes for models of intermediate-mass stars (initial masses correspond to the numbers given in the plots) computed for different metallicities (Z = 3$\times 10^{-4}$, 10$^{-3}$, and 8$\times 10^{-4}$; green circles, blue squares, and red triangles respectively).
The two arrows indicate the impact of mass loss for a couple of specific cases.
Figures from \citet{Ventura13}}
\label{Fig:ONaMgAl_venturaetal13}
\end{figure}

The AGB scenario was extensively developed and discussed in the literature, first on qualitative arguments, and later based on the development of custom-made models of intermediate-mass stars (e.g.\citealt{Ventura01,Ventura02,VenturaDa11MgAlSi,Ventura13,DenissenkovHerwig03,KarakasLattanzio03,Herwig04,Fenneretal04,VenturaDa05a,VenturaDa05b,VenturaDa09,Decressin09a}).
These studies have pointed out several severe difficulties in building the observed chemical patterns from the theoretical yields predicted by TP-AGB models. 
The main problem is due to the competition between i) {\it{hot-bottom burning}} that modifies the envelope abundances via the CNO-cycle and the NeNa- and MgAl-chains,  and ii) {\it{third dredge-up events}} (3DUP) that contaminate the AGB envelope with the helium-burning ashes produced in the thermal pulse and that are actually not observed in GC 2P stars. 

On one hand, 3DUP brings to the stellar surface primary $^{12}$C and $^{16}$O (which leads to an increase of the total C+N+O in the AGB ejecta),
as well as (primary) $^{22}$Ne and $^{25, 26}$Mg produced via successive $\alpha$-captures on $^{14}$N in the convective tongue that develops due to the bolting of triple-$\alpha$ reactions during the thermal pulses. 
This production of the neutron heavy Mg isotopes in the He-shell flash via $^{22}$Ne($\alpha$,n)$^{25}$Mg and $^{22}$Ne($\alpha,\gamma)^{26}$Mg
depends on the abundance of the matter left by the HBS at the end of the interpulse phase. 
It requires temperatures higher than $\sim 3 \times 10^8$~K that are reached typically in the He-burning shell (HeBS) for stars initially more massive than $\sim$3~M$_{\odot}$ and that strongly depend on the reaction rates.
Finally, the Al isotopes are not produced in the HeBS.

On the other hand, if the temperature at the bottom of the convective envelope is high enough (which is typically the case for stars with masses $\geq$ 4~M$_{\odot}$),
HBB modifies the envelope abundances via the CNO-cycle and the NeNa- and MgAl-chains.
HBB results in the production of $^{14}$N, in the depletion of $^{15}$N, $^{18}$O, and, if the temperature is high enough, of $^{16}$O.
When the NeNa- and MgAl-chains operate at higher temperature, $^{23}$Na and $^{26}$Al are produced at the expense of the dredged-up $^{22}$Ne and $^{25}$Mg
respectively (in addition to the important increase of the surface $^{23}$Na abundance that already occurs as a result of the
second dredge-up from the conversion of the initial abundance of $^{22}$Ne and some $^{20}$Ne by H-shell burning; e.g. \citealt{Mowlavi99}).
The initial total Mg in stars is mainly under the form of $^{24}$Mg.
In the case of very high HBB temperatures, proton-captures on $^{23}$Na and $^{24}$Mg deplete the surface abundances of the elements,
and some $^{25, 26}$Mg and $^{27}$Al are produced.

\begin{figure}[h]
\centering
\includegraphics[width=0.5\textwidth]{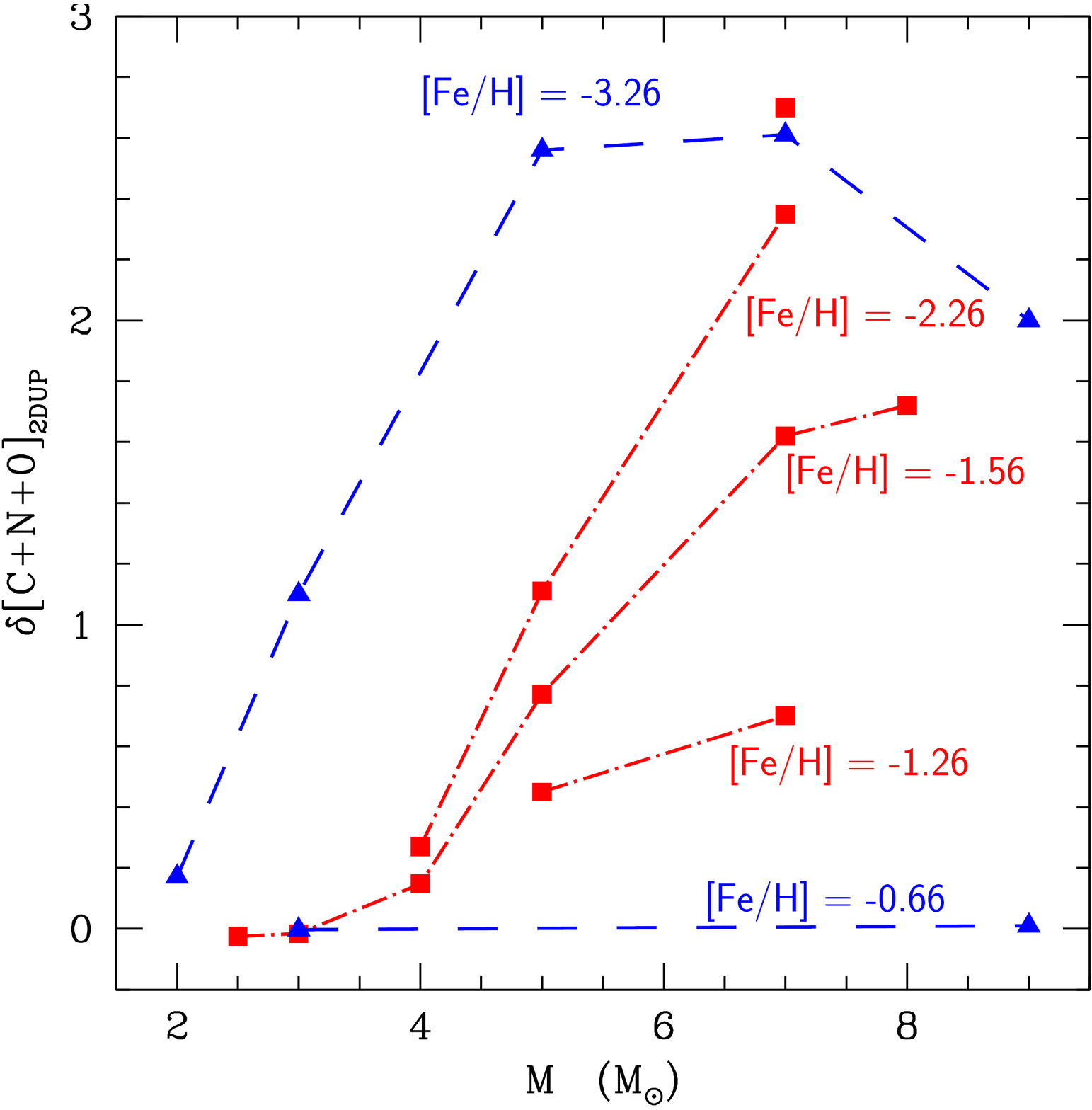}
\caption{Increase of C+N+O expected in rotating stellar models at the end of the second dredge-up on the early-AGB as a function of initial stellar mass and for various values of [Fe/H].
Figure from \citet{Decressin09a}}
\label{Fig:CNOAGBDecressin09a}
\end{figure}

The competition between 3DUP and HBB is crucial for the nucleosynthesis in massive and super-massive AGB stars. 
It strongly depends on metallicity as well as on the assumptions made when building stellar models (e.g. \citealt{Forestini97,LattanzioWood03,Karakasetal06He,WeissFerguson09,Siess07,Siess08,Siess10,Karakas10,MarigoGirardi07,Marigoetal13,Ventura13,Dohertyetal14II,Dohertyetal14III}).
Several important macrophysical processes that have long been identified as major players in AGB evolution are not yet based on solid theoretical developments but rather heavily parametrized in the modeling (e.g. \citealt{HabingOlofsson03,Herwig05}).
While convection settles the beginning as well as the extent of HBB, it is still currently treated in most of the models with the mixing-length theory \citep{Bohm58,Weissetal04}. 
Mixing processes at the interface between the base of the convective envelope and the HBS are far from being understood, although their parametrization dictates 3DUP efficiency (e.g. \citealt{Herwig00}). 
Mass loss, which controls the number of thermal pulses that a star can undergo as well as the maximum HBB temperature and the quantity of stellar ejecta of a given composition, is treated with a simple expression that depends on the stellar luminosity and effective temperature and on an efficiency parameter (e.g. \citealt{Wachteretal02,vanLoonetal05}).
Molecular opacities also play a significant role on the effective temperatures and mass-loss rates of TP-AGB stars that experience 3DUP \citep{Marigo02}. 
Last but not least, some nasty numerical uncertainties can affect the models predictions (e.g. \citealt{Mowlavi99}). 
Severe uncertainties thus plague AGB models,  and we agree with \citet{Renzini15} on the fact that ``stellar models [for AGB] can ÒpredictÓ ...
quantities as functions of various, arbitrary input parameters, but almost certainly they are all wrong as so far only a tiny fraction of the parameter space has been explored". 

As of today however, it is largely recognized that {\it{with the input physics currently considered in the stellar models}} (and which remains perfectible although it has been successful in explaining AGB observations in other astrophysical contexts; e.g. \citealt{Venturaetal15PNe}), {\it{it is not possible to obtain AGB yields presenting simultaneous O depletion and Na enrichment and keeping the C+N+O sum constant as required by the observations in GCs}}. 

Figure~\ref{Fig:ONaMgAl_venturaetal13} presents  O, Na, Mg, and Al yields obtained for AGB models of different initial masses and metallicities \citep{Ventura13}. 
This shows that {\bf{O and Na are correlated in all cases}}.
As clearly stated by \citet{Ventura13} ``{\it{the positive correlation between O and Na is expected independently of all the uncertainties affecting the predictions concerning the Na yield}} (initial Ne and ¥Na abundances, cross-sections of the NeÐNa nucleosynthesis), that may eventually shift upwards or downwards the trend defined
in" Fig.~\ref{Fig:ONaMgAl_venturaetal13} ``without changing the slope." 
In the same figure, one sees also the modifications of Mg and Al due to the combined and opposite effects on Mg due to 3DUP that increases $^{25,26}$Mg and HBB that requires temperatures at the base of the convective envelope of the order of 10$^8$~K to destroy $^{24}$Mg \citep{PCI07,VenturaDa11MgAlSi}. 
The problem is that O and Na are also destroyed at such high temperatures.
As a result, the trends of [Mg/Fe]  and O with  initial stellar mass are similar, and 
there is no possibility to get Na enrichment when O and Mg are depleted.
\citet{Renzinietal15} suggest a reduction of the reaction rate $^{23}$Na(p,$\alpha$)$^{20}$Ne by a factor 5 to reduce the problem of the simultaneous destruction of O and Na in TP-AGB stars; this remains an option, although the necessary change is beyond the reaction rate uncertainties published by \citet{Haleetal04}. We also refer to 
\citet{DenissenkovHerwig03} and \citet{Karakasetal06_22Ne} for a discussion on the uncertainties on the production of heavy Mg isotopes in AGB stars in relation with the uncertainties on the nuclear reaction rates. Importantly, the HBB temperature may be high enough to also affect the abundances of K and Si in the most massive AGB \citep{Ventura12}, but in that case Na is strongly depleted and no Na-K correlation can be obtained theoretically (\citealt{Iliadisetal16}; Prantzos, Charbonnel, \& Iliadis, in prep.), in contradiction with the observations. 

Perhaps one of the strongest nucleosynthetic arguments against AGB stars being GC polluters concerns the predictions for the variations of the sum C+N+O, which increases due to 3DUP. 
Additionally, during central He-burning rotational mixing transports fresh CO-rich material from the core into the HBS, as shown in rotating models of intermediate-mass (2.5 -- 8~M$_{\odot}$) covering the [Fe/H] range of Galactic GC \citep{Decressin09a}. This leads the production of primary $^{14}$N. 
For the most massive AGBs (typically above 3 -- 4~M$_{\odot}$, depending on metallicity; e.g. \citealt{BeckerIben79}), the stellar convective envelope digs this reservoir
during the second dredge-up episode (2DUP) on the early-AGB, which results in a large increase in the total C+N+O at the stellar surface hence in the AGB ejecta (Fig.~\ref{Fig:CNOAGBDecressin09a}). 
This is at odds with C+N+O being constant from star-to-star in GC 2P.

Finally, the 2DUP increases the surface helium abundance of intermediate-mass stars on the early-AGB. Consequently, the yields of massive AGB stars are enriched in He. 
In the models of \citet{Ventura13} shown in Fig.~\ref{Fig:ONaMgAl_venturaetal13}, the helium content of the ejecta increases with mass, and it reaches a value of 0.38 (in mass fraction) for the most massive stars (super-AGBs). This is in total agreement with predictions from other AGB models (e.g. \citealt{Forestini97,Siess10,Dohertyetal14II}).  
Importantly, {\it{the AGB yields in He on one hand, and of O, Na, Mg, Al on the other hand, are not strictly correlated}}, as these elements are processed at different moments of the evolution of the stars, i.e., by 2DUP on one hand, and HBB and 3DUP on the other hand (He enrichment due to HBB is negligible). 
As discussed in \S~\ref{section:2Pevolution}, the resulting He content of 2P stars is an important ingredient to understand their actual evolution and properties.

In summary, {\bf{if a stellar population/generation forms directly from the ejecta of massive  AGBs, it should show (1) a correlation between O and Na, (2) C+N+O enrichment with respect to 1P stars, (3) Mg-depleted stars should be Na-poor, and (4) all 2P stars should have similar and relatively high helium}}. All these theoretical predictions are in clear conflict with the observations of the chemical composition of GC 2P stars.

{\it{et de hoc satis}} with the AGB hypothesis? It is maybe too early to say if one considers only nucleosynthetic arguments, since the modeling of the thermal pulses can certainly be improved with better input physics. However, the C+N+O enrichment due to rotation before the beginning of the TP-phase is certainly a solid result that will be difficult to reconcile with the observed constancy of this quantity within GCs. 
Yet, the global AGB scenario calls for dilution of the gas ejected in the winds of AGB stars with pristine, uncontaminated intracluster gas in order to turn the theoretical ONa-correlation into an anti-correlation, as we shall discuss in \S~\ref{subsection:AGBscenario}. 

\subsection{Fast Rotating Massive Stars - Nucleosynthesis}
\label{subsection:FRMSnucleosythesis}

Following \citet{Norris04} and \citet{Maeder06} who proposed that massive stars could be at the origin of the high helium sequence observed in $\Omega$~Cen, and \citet{Prantzos06} who argued that massive stellar polluters may be a viable alternative to the GC AGB scenario, \citet{Decressin07b} showed that Fast Rotating Massive Stars (FRMS) could be at the origin of GC abundance patterns. 

\begin{figure}
\centering
\includegraphics[width=0.48\textwidth]{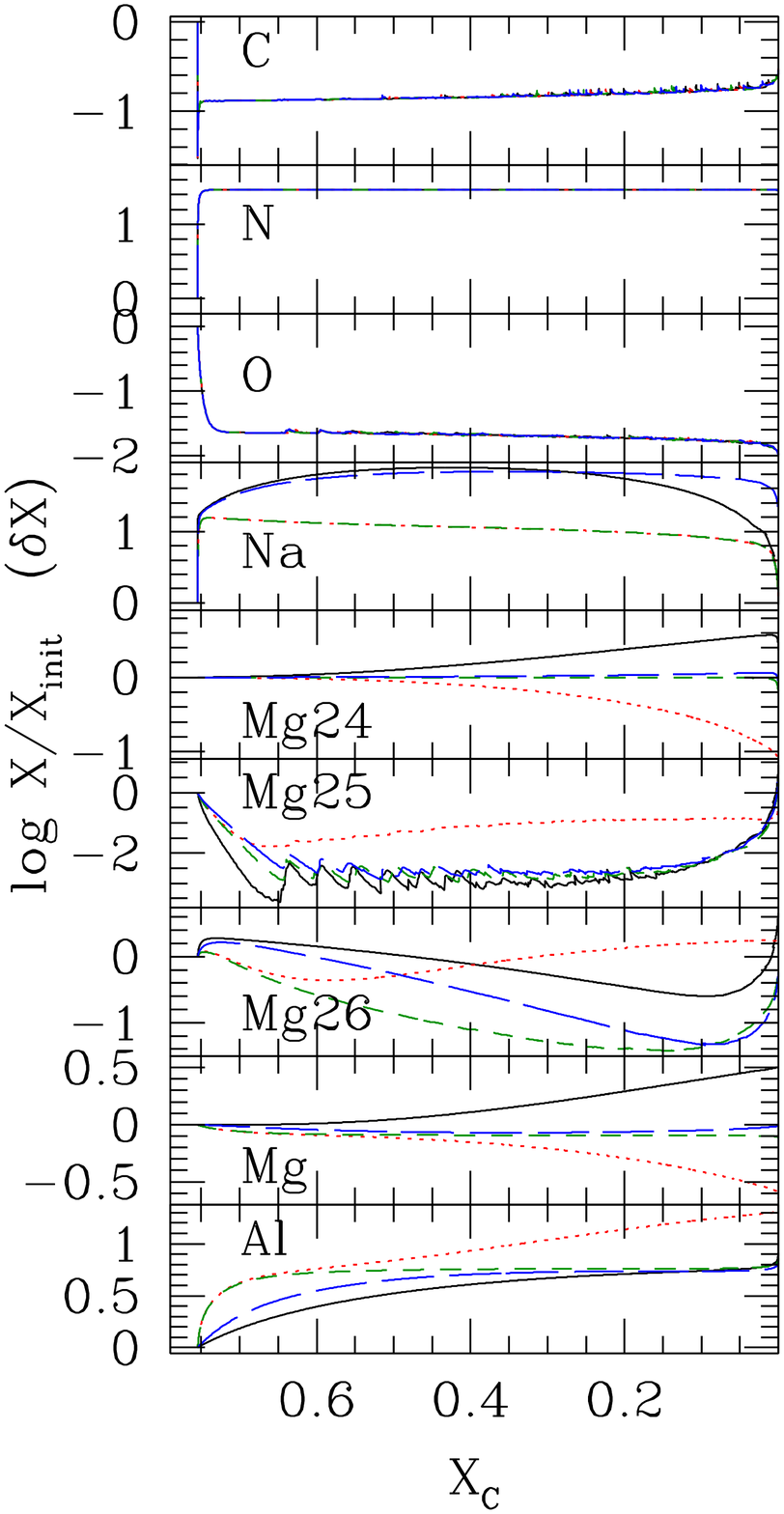}
\includegraphics[width=0.48\textwidth]{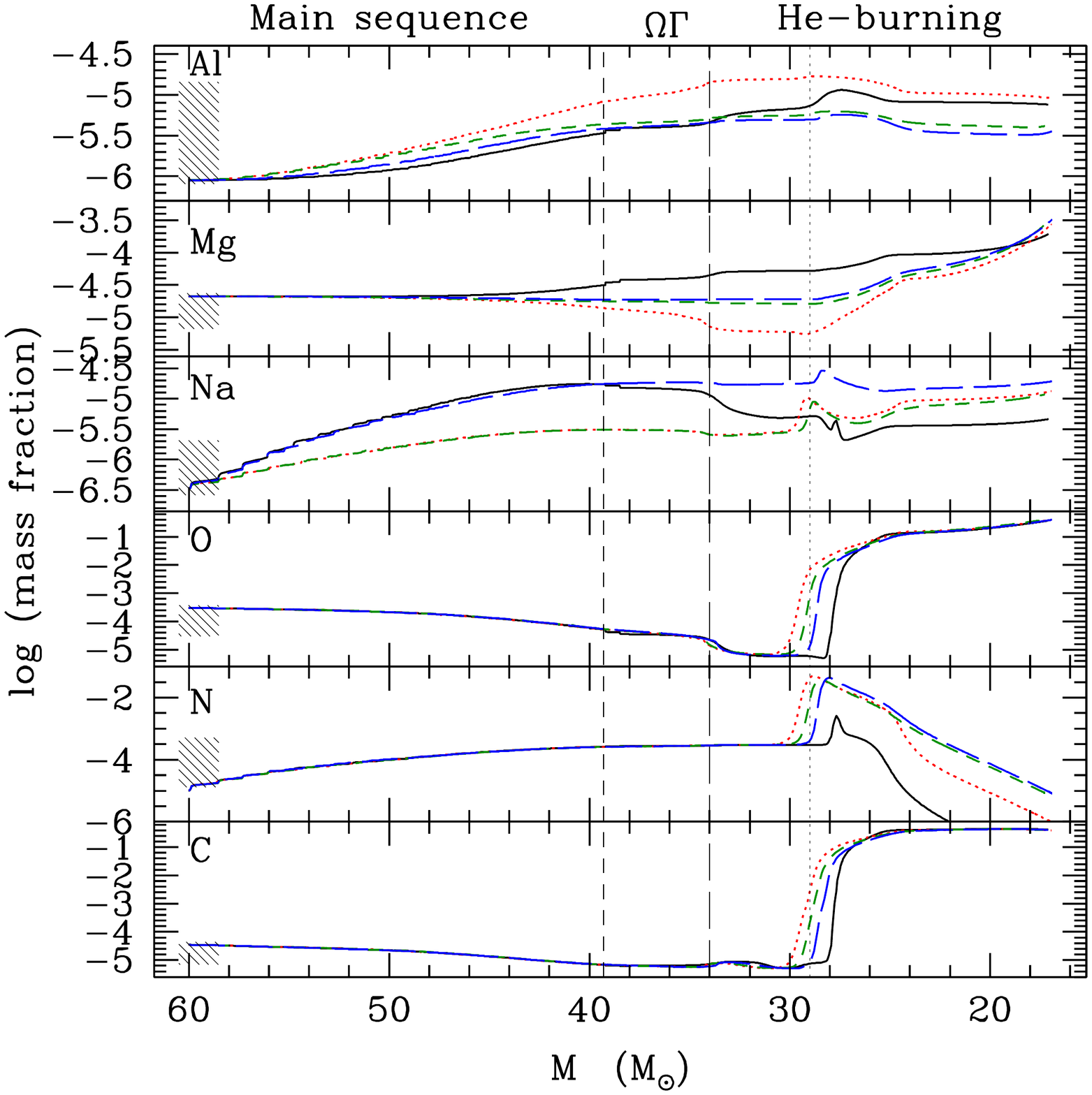}
\caption{Evolution of the central and surface abundances (left and right respectively) in a Fast Rotating Massive Star (initial mass 60~M$_{\odot}$; metallicity Z = 0.005) computed by \citet{Decressin07b}. The abscissa are respectively the H mass fraction at the center of the star (X$_C$; only MS) and the total stellar mass (MS and beginning of central He-burning phase) that both decrease following temporal evolution from left to right.  Different colors refer to different sets of nuclear reaction rates adopted in the model computations (see the original paper for details). The shaded areas in the left panels depict the amplitudes of the abundance variations observed at the surface of low-mass stars in the GC NGC~6752.
Figures from \citet{Decressin07b} }
\label{Fig:Detal07_AbundCen_AbundSurf}
\end{figure}

The central temperatures of massive main-sequence stars (i.e., $\geq$ 20 -- 25~M$_{\odot}$) reach the high values required to produce most of the observed abundance anomalies (Fig.~\ref{Fig:PrantzosCI06fig6}).
In the deep interior of these objects, the CNO-cycle is at equilibrium and the NeNa-cycle is efficient already very early on the main sequence. Therefore, the O-Na and C-N anticorrelations built up easily, the extent of Na-production depending on the adopted reaction rates.  Additionaly, Al is produced.
The only difficulty concerns Mg, which is hardly predicted to decrease with the currently available nuclear reaction rates, and which comes at the end of the main sequence, thus with non-negligible helium enhancement (see below). In order to account for $^{24}$Mg depletion, \citet{Decressin07b} increased the rate of the reaction $^{24}$Mg(p,$\gamma$) reaction by a factor of 1000 with respect to the nominal value of \citet{Iliadisetal01}. In that case, the Mg-Al anti-correlation also builds up in the stellar core ($^{24}$Mg decreases, $^{26}$Mg increases, and $^{25}$Mg remains approximately constant). 
Predictions abundances at the center of a 60~M$_{\odot}$ FRMS ¥model with metallicity Z = 0.005 are shown in Fig.~\ref{Fig:Detal07_AbundCen_AbundSurf} \citep{Decressin07b}.
These theoretical abundance patterns are in agreement with the general chemical trends observed in GCs.

\citet{Decressin07b} proposed that fast rotation was responsible for the enrichment of the whole interior as well as the winds of FRMS with the H-burning products, due to meridional rotation and hydrodynamical instabilities that transport nuclides from the stellar center to the surface (see e.g. \citealt{Maeder09,MaederMeynet12,Palacios13} for reviews on rotation-induced processes in stars and their impact on stellar evolution and nucleosynthesis).
Additionally, high rotational velocity strongly affects stellar mass loss. When a star rotates at break-up velocity, it is expected to lose a very important fraction of its initial mass
through a mechanical wind that has a very low velocity (hence can be retained within the cluster). The expelled matter very likely forms an equatorial disk around the star, similar to that observed around Be-type stars (e.g. \citealt{Riviniusetal01,PorterRivinius03,Townsendetal04,Hauboisetal12,Granadaetal13}). 
The predictions for the resulting surface abundances of C, N, O, Mg, and Al are shown in Fig.~\ref{Fig:Detal07_AbundCen_AbundSurf} for a 60~M$_{\odot}$ FRMS model with metallicity Z = 0.005 \citep{Decressin07b}. In this specific model, mass loss during the whole main sequence and the beginning of the central He-burning phase at the $\Omega \Gamma$-limit\footnote{\citet{MeynetMaeder00} define three kinds of break-up limits depending on the mechanism that counterbalances gravity and leads to break up: Rotation ($\Omega$-limit), radiation ($\Gamma$-limit), rotation and radiation ($\Omega \Gamma$-limit).} 
is dominated by this strong rotation-induced mechanical wind and is strongly increased with respect to the standard case. During that period, the 60~M$_{\odot}$ FRMS model loses 40~M$_{\odot}$ (to compare with a 60~M$_{\odot}$ model of similar Z that loses less than 1~M$_{\odot}$ through classical radiation-driven winds). When the star moves away from the $\Omega \Gamma$-limit due to heavy mass loss, the fast radiatively-driven winds take over;  the equatorial ``decretion"-disk is disconnected from the star so that it is cannot be fed by the He-burning ashes (see the corresponding strong increase in C and O when  the remaining mass of the 60~M$_{\odot}$ model is $\sim$ 28~M$_{\odot}$) nor by the products of the subsequent nuclear burning phases. 
During the whole evolution phase when FRMS rotate at or near break-up and matter is ejected by gentle blowing winds that can  be  easily  retained  within  the  GC  potential  well, the surface abundance variations do mimic the central ones with some delay due to non-instantaneous rotational mixing (Fig.~\ref{Fig:Detal07_AbundCen_AbundSurf}). 
Importantly, the material ejected by massive stars is lithium-free, as Li is a very fragile element (it burns at $\sim 2.5$~MK in stellar interiors). Therefore, {\bf {dilution of slow winds with pristine proto-cluster material in the immediate surrounding of individual massive stars}} is necessary to explain the presence of Li in 2P stars. Actually, the  observed Li-Na anticorrelation in GCs (Fig.~\ref{Fig:Li_NGC6397_Lind2009}) can be directly used to determine the dilution ratio between the slow wind and interstellar material necessary to form 2P stars (e.g. \citealt{Decressin07a,Lindetal09,Chantereau16}).

In summary, in the original FRMS models {\bf{the C-N and O-Na anticorrelations as well as the Al increase appear naturally in the FRMS yields}}. An important increase of the rate of the reaction $^{24}$Mg(p,$\gamma$) is however necessary to produce a Mg-Al anticorrelation. 
Additionally since helium is the main outcome of H-burning the {\bf{abundance patterns are predicted to be correlated to various degrees with helium enrichment}}\footnote{This can be seen easily in  Fig.~\ref{Fig:Detal07_AbundCen_AbundSurf} where the evolution of the central abundances is shown as a function of the remaining hydrogen mass fraction X in the core, which decreases as the helium mass fraction Y increases (X+Y+Z=1, where Z=0.0005 does not change along the MS).}. So far, the theoretical He enrichment is higher than allowed by observations and by isochrone fitting \citep{Bastianetal15,Chantereau16} as discussed in details in \S~\ref{section:2Pevolution}.

{\it{et de hoc satis}} with the FRMS hypothesis? It may be too early to say (as for the AGB scenario), since the input physics of the FRMS models is also highly perfectible.
In particular, the prescription for the treatment of mass loss at critical velocity that was used by \citet{Decressin07b} implies that the ejection of the H-burning ashes through slow equatorial winds proceeds all along the MS evolution and the luminous blue variable phase.  This results in a correlation between helium and sodium content in 2P stars  \citep{Chantereau15,Chantereau16}.
However, central temperatures in stars more massive than $\sim$ 25~M$_{ \odot}$ become high enough for the NeNa-chain to operate very early on the MS, and $\sim 60 - 70 \%$ of the stellar mass that is included in a very extended convective core reaches very quickly the correct chemical composition to explain the Na without too strong He enrichment. Therefore, if some additional mechanism (e.g. stellar pulsation, hydrodynamical instabilities, ...) could increase the mechanical mass loss at critical rotation, the problem of too much He enrichment could be overcome. However, the Mg problem would remain if the nuclear reaction rates and the constraints on helium enrichment had to be confirmed. Indeed, even in the most massive stars considered in the original FRMS scenario, i.e., 120~M$_{\odot}$, central temperatures as high as $\sim 75$~MK required to decrease $^{24}$Mg are reached only in the second half of the MS, and the same is true for 200 and 500~M$_{\odot}$ stars (C.Georgy, private communication) and for $10^3$~M$_{\odot}$ stars \citep{DenissenkovHartwick14}. In this context, super-massive stars with masses of the order of $10^4$~M$_{\odot}$ might be the best candidate polluters as discussed below.

\subsection{Supermassive stars - Nucleosynthesis}
\label{subsection:Supermassivestarsnucleosythesis}

\begin{figure}
\centering
\includegraphics[width=\textwidth]{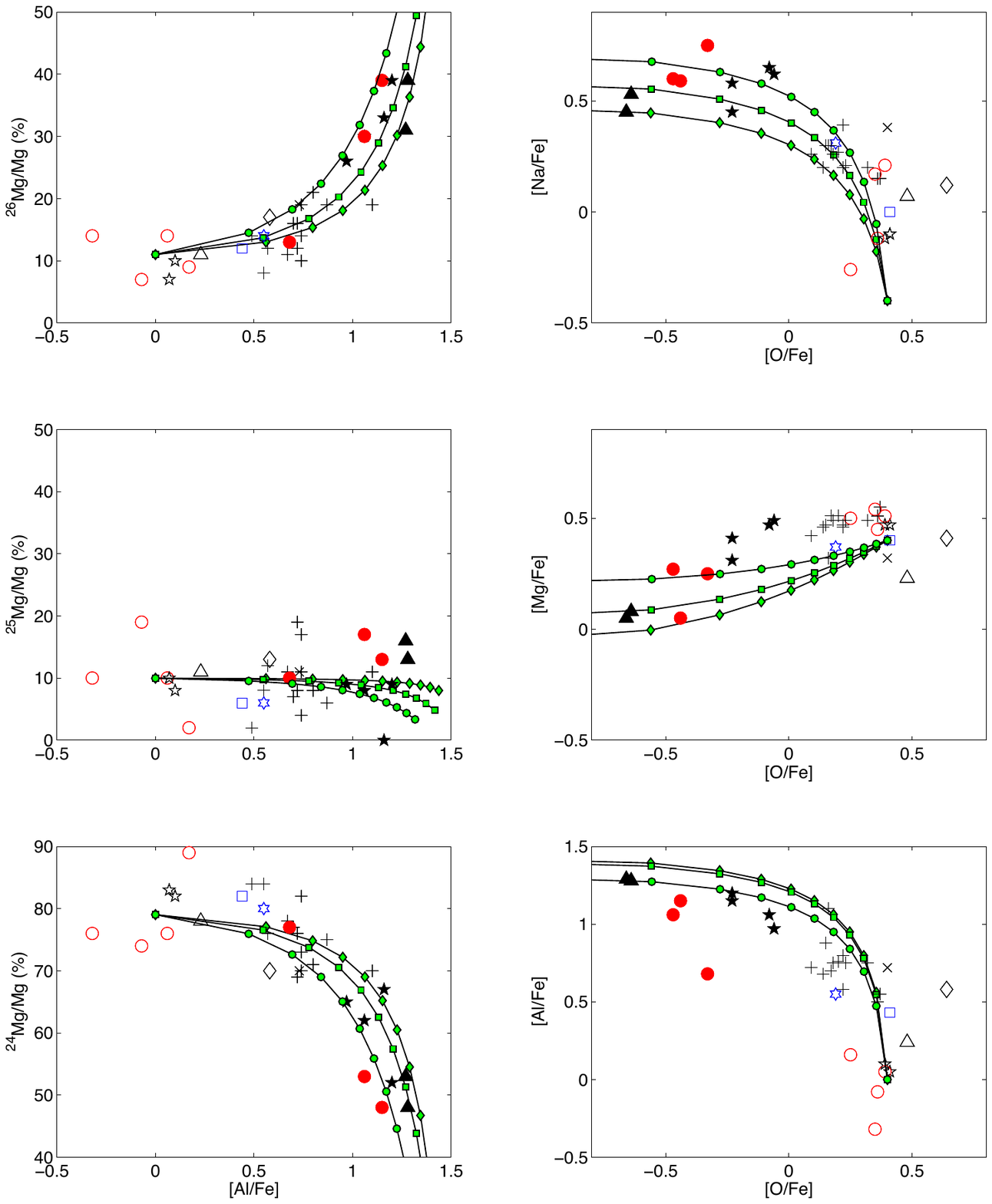}
\caption{Predictions for the composition of the mixture composed of SMS ejecta (assuming the composition of the H-burning ashes at the time on the MS when the central helium mass fraction Y has increased from 0.25 to 0.4) diluted with pristine gas. Dilution varies from 1 to 0 with 0.1 steps along the theoretical lines; green symbols connects predictions for SMS of 2, 3, and 4 $\times 10^4$~M$_{\odot}$ (circles, squares, and diamonds respectively). The other symbols are observations for stars in different clusters ($\Omega$~Cen, M~13, M~71, and M~4).
Figures from \citet{DenissenkovHartwick14} }
\label{Fig:DH14_SMSnucleosynthesis}
\end{figure}

\citet{DenissenkovHartwick14} proposed supermassive stars (SMS, with masses $\sim 10^4$~M$_{\odot}$) as the main source of H-burning products for GC self-enrichment. 
They assume that SMSs form in GCs either via runaway collisions of massive stars at the cluster center  \citep{PortegiesZwartetal04,Freitagetal06}, or directly via the monolithic collapse of the proto-cluster cloud. Hypothetical SMSs are fully convective, they are dominated by radiation-pressure and have super-Eddington luminosities ($10^9 - 10^{10}$~L$_{ \odot}$) with very high effective temperature (of the order of $10^5$~K). According to the models of Denissenkov, these objects reach central temperature for CNO, NeNa, and MgAl already at the beginning of the main sequence. The authors thus arbitrarily postulate incomplete H-core burning: In order to restrict the He abundance increase to the isochrone-fitting constraints, they assume that central H-burning occurs only for a very limited fraction of the main sequence, and ends when the He mass fraction Y has increased by $\sim$0.15 only (i.e., Y$\sim$0.4); this occurs in a few $10^5$ yrs. 
They invoke SMS fragmentation by diffusive mode of the Jeans instability \citep{Thompson08}  or extreme mass-loss due to super-Eddington radiation continuum-driven wind  \citep{DotanShaviv12} to make SMS release most of their material at this specific evolution moment, while the remaining stellar inner regions would eventually collapse and form intermediate-mass black holes. 

In this framework, 2P stars are assumed to form from SMS H-burning ashes mixed with pristine gas to various degrees. Dilution curves are shown in Fig.~\ref{Fig:DH14_SMSnucleosynthesis} for the yields of SMS with different initial masses (2, 3, and 4 $\times 10^4$~M$_{\odot}$). The green symbols along the theoretical lines correspond to different mixtures made of a {\it{(1-d)}} part of SMS material (as predicted by the SMS models exactly when central He mass fraction Y equals 0.40) and an {\it{d}} part of pristine GC gas; the dilution factor  {\it{d}} decreases from 1 to 0 starting from the composition of 1P stars (the right extremity of the
O-Na anticorrelation; the other extremity corresponds to SMS pure ejecta). The comparison with observational abundance determinations of different isotopes in GC stars is relatively satisfactory. 

In view of these exploratory studies, SMS can be considered as very promising candidates for GC self-enrichment, at least from the nucleosynthesis point of view. However, many open questions remain about the formation and evolution of these hypothetical stars, and a global scenario is still missing (\S~\ref{subsection:SMSscenario}).

\section{Global self-enrichment scenarios}
\label{section:selfenrichmentscenario}

\subsection{Global AGB scenario}
\label{subsection:AGBscenario}

Despite the contradictions between spectroscopic observations in GCs and nucleosynthetic predictions of AGB models computed by different research groups (\S~\ref{subsection:AGBnucleosythesis}), a complex AGB scenario has been developed \citep{VenturaDa05c,D'ercole08,D'ercole10,D'ercole11,D'ercole12,Bekki11,D'Antonaetal16}. 
In order to ``save" (quoting the authors, see e.g.  \citealt{Renzinietal15}) the AGBs as the source for GC self-enrichment,  it invokes in particular {\it{dilution of the gas ejected in the winds of AGB stars with pristine, uncontaminated intracluster gas in order to turn the theoretical ONa-correlation into an anti-correlation}}.  

In the model developed by \citet{D'ercole08}, GCs undergo a first episode of star formation, where 1P stars form out of pristine proto-cluster gas (i.e., with the chemical properties of halo stars). Then supernova feedback is invoked to expel all the remaining initial gas as well as the SNe ejecta.  Gas expulsion is also expected to change the cluster potential well so that most 1P low-mass stars can be ejected (see the more general discussion on the mass budget problem in \S~\ref{subsection:massbudget}). 
After the end of the SNe phase, the yields of massive and supermassive AGB stars accumulate in the cluster center where the escape velocity is higher than AGB wind velocity. This establishes a cooling flow towards the cluster center where AGB yields accumulate and are available to give birth to 2P stars. Accretion of fresh gas is then required for dilution with AGB ejecta in order to produce the O-Na anticorrelation among 2P stars. Possible successive bursts of 2P star formation separated by periods of star formation inactivity are expected to lead to discreetness between 2P stars as depicted by photometry (e.g. \citealt{D'Antonaetal16}). 

If we follow this temporal sequence of events, we can identify a series of difficulties encountered by this model (in addition to the nucleosynthesis issues discussed in \S~\ref{subsection:AGBnucleosythesis}). 

One issue concerns gas ejection by SNe explosions that is required to expel the pristine gas left after the formation of 1P stars together with Type II SNe ejecta, and that is invoked to unbind the vast majority of 1P low-mass stars so as to solve the mass budget problem \citep{Decressin07a,Bekkietal07,D'ercole08,SchaererCharbonnel2011}. For this to work, N-body simulations require explosive gas expulsion on the crossing timescale of the cluster \citep{Decressin10b,KhalajBaumgardt15}.
Stellar winds and supernovae explosions fail however to produce sufficient power for this process to work, as shown in 
gas expulsion models resulting from the formation of a superbubble (with kinematics described by a thin-shell
model; \citealt{Krauseetal12,Krauseetal13}). With these sources of energy indeed, the superbubbles are destroyed by the Rayleigh-Taylor instability 
before they reach escape speed for all but perhaps the least massive and most extended clusters. More issues related to the key problem of gas expulsion are discussed in \S~\ref{subsection:massbudget}.

Another issue concerns the duration of the formation of 2P stars from AGB stars that can in principle last between  $\sim$ 40 and 200~Myrs, depending on the initial mass range of the potential AGB polluters, and ending when Type Ia SNe start exploding. This is in tension with observations of a large sample of extragalactic Young Massive Star Clusters (YMC) which ages fall into this domain and which masses and compactness are similar to those of GCs (e.g. \citealt{Larsenetal11,Bastianetal2013b,CabreraZirietal2015,Niederhoferetal15}). Indeed, none of these YMCs were found to show ongoing star formation, nor evidence for multiple bursts of star-formation or of gas or dust retention that is required to form 2P stars. We refer to the lecture by N.Bastian (this volume) for more details on the constraints on GC evolution than can be gathered from observations of YMCs . 

Finally, one of the main difficulties of the AGB scenario is the required re-accretion of gas with the same exact composition as the original proto-cluster. Where does this gas come from? Has gas from the original proto-cluster been stored somewhere, where, and how did it avoid SNe contamination? Or did the cluster accrete fresh gas encountered on its orbit $\sim$ 100~Myrs after its formation? In that case, what are the chances to accrete material with exactly the same content in Fe-peak elements than the original proto-cluster cloud? Clearly, much work remains to be done to answer these questions.

\subsection{Global FRMS scenario}
\label{subsection:FRMSscenario}

\citet{Krauseetal12,Krauseetal13,Krauseetal16} revisited \citet{Decressin07a,Decressin07b} FRMS scenario for the case of ``typical" GCs like NGC~6752 where no star-to-star [Fe/H] abundance variations is observed. They included the constraints from N-body simulations \citep{Decressin10b,KhalajBaumgardt15} and took into account the dynamics of interstellar bubbles produced by stellar winds and supernovae as well as the interactions of FRMS with the ICM and followed the kinematics with a thin-shell model. 
Because of the mass-budget issues summarized in \S~\ref{subsection:AGBscenario}, the total initial mass and half-mass radius of the proto-cluster were assumed to be 9 $\times$10$^6$~M$_{\odot}$ and $\sim$ 3 pc respectively, with a star formation efficiency (SFE) of 30~\%. For a Salpeter stellar IMF this implies that the cluster has hosted $\sim$ 5700 1P massive stars (with masses above 25~M$_{\odot}$). 
The cluster is therefore strongly impacted by the fast radiative winds of the numerous massive stars during the very first Myrs after the formation of the 1P.
However, those winds are inefficient in lifting any significant amount of gas out of the GC because their integrated energy is much smaller than the binding energy of the intracluster material (ICM).  They are nevertheless expected to create large hot bubbles around massive stars, and to compress the ICM into thin filaments.  ``Classical" star formation is inhibited in such an environment by the high Lyman-Werner flux (see also \citealt{ConroySpergel11,SchaererCharbonnel2011}) during the wind bubble phase as well as the subsequent supernova phase. 

However, if massive stars have equatorial mass ejections that form decretion disks as originally proposed by \citet{Decressin07b} and as expected in the case of fast rotation, 
accretion of pristine gas may proceed in the shadow of the equatorial stellar ejecta if the orbits of the FRMS are sufficiently eccentric.
Massive discs around individual polluters are then fed both by pristine gas from the outside and by H-burning ashes ejected by the FRMS from the inside. 
The total amount of gas that is released and can potentially form 2P stars through this mode and with abundance properties similar to those observed today is high. 
Indeed \citet{Decressin07b} models predict that FRMS lose about half of their initial mass via equatorial ejections loaded with H-burning products during the main sequence and the luminous variable phase.
As a consequence, the total mass lost through this mechanism by all the FRMS is of the order of 10$^4$~M$_{\odot}$ per Myr. 
On the other hand, a similar amount of pristine gas is expected to be brought by the accretion flow, with mixing proportions depending on the FRMS orbit-averaged accretion rate. 
2P stars are then expected to form sequentially through gravitational instability in their parent FRMS disc with the current light element abundances of the respective disc;  the increased pressure from the SN may also compress its associated disc such that the last 2P stars form at this occasion. 
 {\it{Overall star formation is expected to be complete after between $\sim$ 3 and 8~Myrs}}, 
the exact duration depending on the mass limit (and therefore the lifetime) at which stars explode as SNe or turn 
silently into black holes, which value is currently highly uncertain. 
Indeed while stars initially less massive than 25~M$_{\odot}$ are expected to produce energetic SNe-events or, eventually, gamma-ray bursts for fast rotating progenitors (e.g. \citealt{Yoonetal06,Dessartetal12}), the situation is much less clear for the more
massive stars that may rather turn silently into black holes (e.g. \citealt{Fryer99,Belczynskietal12}).

As for the case of the AGB scenario, the FRMS scenario encounters several difficulties and makes assumptions that need to be assessed (in addition to the nucleosynthesis issues discussed in \S~\ref{subsection:FRMSnucleosythesis}). 

As discussed extensively in \citet{Krauseetal13}, an important issue concerns the need for efficient accretion of pristine gas onto the FRMS discs. This requires eccentric orbits 
of the massive stars, with low angular momenta and velocities near the outer turning points below about 10 $\%$ of  the circular velocity. Whether this could be specific to GC formation scenario remains to be understood, and potentially tested observationally. 

The phenomena of equatorial ejections by fast rotating stars is known from Be-type stars (references above). 
Models are developed to understand the spreading out of such discs in terms of viscous evolution similarly to what \citet{Krauseetal13} propose for discs around FRMS (e.g. \citealt{Hauboisetal12}).
However, formation of self-gravitating bodies in discs is still an open question. It has recently begun to be explored, first in the context of planet formation (e.g. \citealt{KleyNelson12}, and references therein), and also in the context of the formation of the first massive stars that may also have been very fast rotators \citep{Clarketal11,Stacyetal11,Greifetal12}. This certainly needs to be worked out through dedicated numerical simulations.

Finally, the gas expulsion and mass budget issues that are common to all self-enrichment scenarios are discussed in more details in \S~\ref{subsection:massbudget}.

\subsection{Supermassive stars - The need for validation of the physics and for a global scenario}
\label{subsection:SMSscenario}
Supermassive stars appear to be extremely attractive from the nucleosynthesis point of view, in particular because they have the requested H-burning temperature to reproduce the Al and Mg abundance variations that have been found in some GCs (\S~\ref{subsection:Supermassivestarsnucleosythesis}). However, several assumptions are made about the formation and the evolution of these (still) hypothetic stars, and no global scenario for GC enrichment by SMS exists so far. Many open questions remain.
How do these SMS stars form, and is their formation specific to GC environments? How many form per GC? How are these stars disrupted, and why does disruption or heavy mass loss happen at a very specific time on the MS (i.e., for a precise He enrichment)? How much of their ashes is made available for 1P star formation, and how much remains trapped into the stellar remnant (possibly an intermediate-mass black hole)? How do their ejecta mix with pristine gas (and is there pristine gas available)? Which mechanism induces the secondary star formation episode? This clearly opens interesting avenues in the domain and extraordinary exploration territories for stellar physics in the near future.

\subsection{Mass budget and gas expulsion}
\label{subsection:massbudget}

When considering a classical stellar IMF for the polluters and accounting for the material that can be potentially made available by 1P polluters to form 2P stars, both the FRMS and the AGB scenarii face the so-called {\bf{``mass-budget" problem}} \citep{Prantzos06,Decressin07b,D'ercole08,Decressin10b,Carrettaetal10,SchaererCharbonnel2011}. 
When assuming that the maximum amount of H-processed ejecta are entirely recycled into the second stellar population after a 50-50\% dilution with pristine gas \citep{Decressin07a}, and that the 2P was composed of low-mass stars only, 
the ratio between 1P and 2P stars is predicted to be of the order of ten to one.
However, the percentage of 2P GC stars observed today in GCs is $\sim 70 \%$, with only slight variations from cluster to cluster (e.g. \citealt{Prantzos06,Carrettaetal13}). 
Therefore {\bf{in the case of a universal stellar IMF, proto-GC must have lost $\sim 95 \%$ of 1P stars in order to explain the present-day 1P to 2P ratio}} \citep{Prantzos06,Decressin10b,D'ercole10,SchaererCharbonnel2011,Krauseetal13,KhalajBaumgardt15}.

\citet{SchaererCharbonnel2011} summarized the situation by highlighting the following key points: \\
(1) ``The initial stellar masses of GCs must have been $\sim$ 8--10 times larger than the current (observed) mass,
in the case where no 2P stars were lost from the GCs. However, if 2P halo stars originated from the present population of GCs, as suggested by observations finding stars characteristics of the 2P in the Milky Way halo \citep{Carrettaetal10,Martell10,Martelletal16}, the initial cluster masses must have been $\approx 18 - 25$ times larger than the current masses." \\
(2) ``The mass in low-mass stars ejected from GCs must be $\sim 2.5-3.2$ times their presently observed 
value if all 2P stars were retained by the GC, or $\sim$ 5--10 times the present day mass if 2P stars were lost.
These numbers translate to a contribution of  5--8\%  or 10--20\% respectively of the ejected low-mass stars to the Galactic stellar halo mass." \\
(3) ``The observations of 2P stars in the Galactic halo can also be used to constrain the initial mass function of the GC population (GCIMF). 
In particular the often assumed power-law with a slope $\beta \approx -2$ is in contradiction with determinations of the fraction of  2P stars in the halo, whereas
a log-normal GCIMF is compatible with these observations. This finding revived the question about a common mass function and about the physical
processes leading to a distinction between GCs with multiple stellar populations and other clusters." \\
(4)  ``Due to their high initial masses, the amount of Lyman continuum photons emitted
by GCs during their youth must have been substantial, reinforcing the conclusion of \citet{Ricotti02} that GCs should have  significantly contributed to reionise the IGM at high redshift (above $\sim$ 6)."

As of today, the most commonly invoked solution to the mass budget issue calls for fast and drastic {\bf{gas expulsion}}: By inducing an important decrease of the cluster's potential well, this is expected to lead to the loss of the vast majority of 1P low-mass stars from the cluster periphery while 2P stars formed in the center of GCs would remain bound to the cluster.  
\citet{Krauseetal12} showed however that gas expulsion via SNe explosions does not work for typical GCs (see also \S~\ref{subsection:AGBscenario}). 
Indeed, even in the cases where the energy released by SNe exceeds the cluster's binding energy, SNe-driven shells are expected to be destroyed by the
Rayleigh-Taylor instability before they reach escape speed. The shell fragments that contain the gas thus remain bound to the cluster. 
Krause and collaborators proposed instead that fast gas expulsion and 1P loss occur thanks to the energy released by the coherent onset of accretion onto the stellar remnants of 1P massive stars (neutron stars and black holes) that is expected happen at the end of the SNe phase when turbulence decreases within the ISM and all massive stars have turned into dark remnants. 
Importantly, \citet{Krauseetal12}  estimated a limiting initial cloud mass of $\sim 10^7$~M$_{\odot}$ for this dark remnant scenario to work. Above that mass, the cold
pristine gas may not be ejected, and subsequent star formation may proceed in the ICM that would then be enriched in SNe ejecta. This could explain the fact that the more massive GCs (Omega Cen, M22, M54, NGC 1851, NGC 3201) do exhibit internal [Fe/H] spread, while typical and also less massive ones do not.

One possible difficulty is that fast gas expulsion through dark remnant accretion is expected to happen $\sim$ 30 -- 40~Myrs after cluster formation. This delay is in tension with observations of young massive star clusters (YMC) in the Local Group and in dwarf galaxies that are used to test the conditions for gas expulsion \citep{Bastianetal14,CabreraZirietal2015,Hollyhead15}. 
Indeed, some of these objects with ages below 10~Myr, current masses between 8 and 50$\times 10^5$~M$_{\odot}$, and half-mass radii between 1.5 and 18~pc (i.e., very similar to the expected properties of GC progenitors), appear to be gas free by an age of $\sim$~3~Myr, and they show no evidence for multiple epochs of star-formation.  
\citet{Krauseetal16} thus concluded that in the case where gas expulsion/depletion proceeds the same way in YMCs than it did in infant GCs, it cannot be obtained by radiation pressure, stellar winds,  supernovae, or  dark remnant scenario as energy sources for fast gas removal when assuming global standard assumptions including classical IMF (\citealt{Salpeter55} or \citealt{Kroupaetal13}) and SFE of 30 $\%$. 
 {\it{For very early gas expulsion to eventually work, one must then invoke either energy ejection by $10^{52}$ - $10^{53}$ erg hypernovae, or a very high star SFE with values ranging from 50 \% in the most loose clusters  up to more than 90 \% in the most compact ones, the efficiency of gas expulsion being extremely dependent on the cluster compactness}} (see details in \citealt{Krauseetal16}). 
However, if gas is cleared simply because of very high SFE then the total stellar mass at the end of the process would not be changed significantly enough to affect the 1P-to-2P ratio as requested by the mass budget. 
Therefore,  {\bf{if GC formation at high redshift was similar to YMC formation happening in the local and modern universe, then gas and star expulsion 
is a serious issue for the self-enrichment scenarios that make classical assumptions for the SFE and for the stellar IMF of the first and second stellar populations.}}

To solve this problem, one can alternatively call for a non-standard, top-heavy IMF for the polluters \citep{SmithNorris82,D'AntonaCaloi04,Prantzos06,SchaererCharbonnel2011}, or for a 1P composed of massive stars only giving birth to all the low-mass stars we observe today in GCs \citep{Charbonnel14}. 
In the case of SMSs, \citet{Denissenkovetal15} circumvent the mass budget problem by assuming that if some pristine gas is available, SMS stars might accrete it for a few $10^6$~years through an accretion disk while expelling H ashes through strong  wind at their poles. If a stable accretion-wind state can be reached by SMS when they have acquired their maximum mass and can last for a few $10^5$ yrs, then SMS can potentially process as much as the present-day GC typical masses (i.e., few $10^5$~M$_{\odot}$). This is however highly speculative as almost nothing is known about the formation and evolution of these hypothetic SMSs.

While these options might open extremely interesting avenues, their many implications have not yet been investigated in detail. On the other hand, there is no definitive evidence yet that GC formation at high redshift was similar to YMC formation happening in the local and modern universe. From the observational point of view, direct relationship between YMCs and GCs could be proven by the detection of the O-Na anticorrelation in YMCs. 
On the theoretical side, relatively simple (i.e., analytical and semi-analytical) modeling of the growth of self-gravitating superbubbles including for the first time relevant prescriptions from stellar models to describe stellar mass loss, winds, luminosity, and energy, have highlighted the very important role of stellar feedback on MSC dynamics (\citealt{Krauseetal12,Krauseetal13,Krauseetal16}, and references therein).  
Detailed aspects of the problem of the formation of stellar populations in the early phases of the evolution of massive star clusters (both YMCs and GCs' progenitors) need to be investigated now with 3D-modeling of all the subtle interactions between stellar ejecta (mass, energy, nucleosynthesis products), interstellar matter, and secondary
star formation.

\section{Impact of their initial composition on the evolution of 2P stars}
\label{section:2Pevolution}

\subsection{Helium: Why do we care?}
\label{subsection:Hewhydowecare}

Old GCs contain today only low-mass stars (below $\sim$ 0.9~M$_{\odot}$) and the dark remnants of the upper end of the initial stellar mass function. 
It is generally well accepted that 2P GC stars have started their life with a higher helium content than their 1P counterparts, since they form from H-burning ashes mixed with pristine gas. This is supported by several observational pieces of evidence (\S~\ref{obshelium}), although the amount of He enrichment is not well constrained yet. 
On the other hand, predictions for the extent of helium enrichment along the sodium distribution in 2P stars strongly vary from one self-enrichment scenario to another, as they depend on the nature of the invoked 1P polluters.
In the AGB scenario, all 2P stars (i.e., spanning a large range of Na abundances) are  expected to be born with very similar helium contents -- maximum of 0.36 - 0.38 in mass fraction (e.g. \citealt{Dohertyetal14II}) if no dilution with the ISM matter is taken into account (to be  compared with $\sim$ 0.248 for 1P stars). 
In contrast, the FRMS scenario in its original form predicts broad and correlated spreads of both helium and sodium in the initial mixture that formed 2P stars, with initial helium mass fraction ranging between 0.248 (as for the 1P) and 0.8 in the most extreme 2P stars \citep{Decressin07b}.
This has noticeable consequences on the evolution and fate of GC 2P stars, that can be used to constrain the self-enrichment scenarios.
Therefore, it is mandatory to account for non-standard helium content on the evolution of low-mass stars when comparing theoretical models with all the spectroscopic and photometric observational constraints, since it definitively impacts their evolution, lifetime, and fate. 

\subsection{Impact of the initial helium content on the evolution of a low-mass star}
\label{subsection:Heenrichment0p8Msun}

\begin{figure}
\centering
\includegraphics[width=0.4\textwidth]{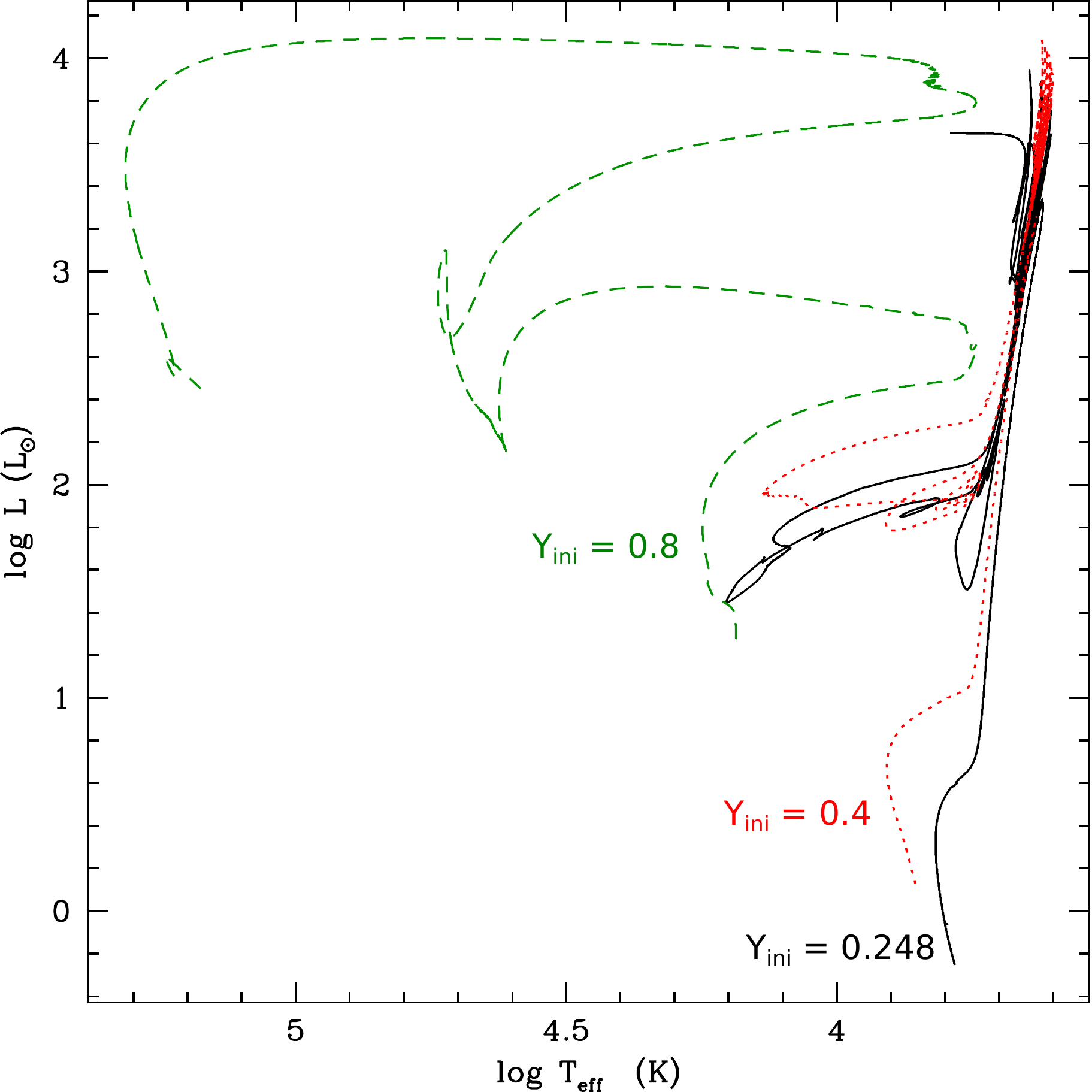}
\includegraphics[width=0.4\textwidth]{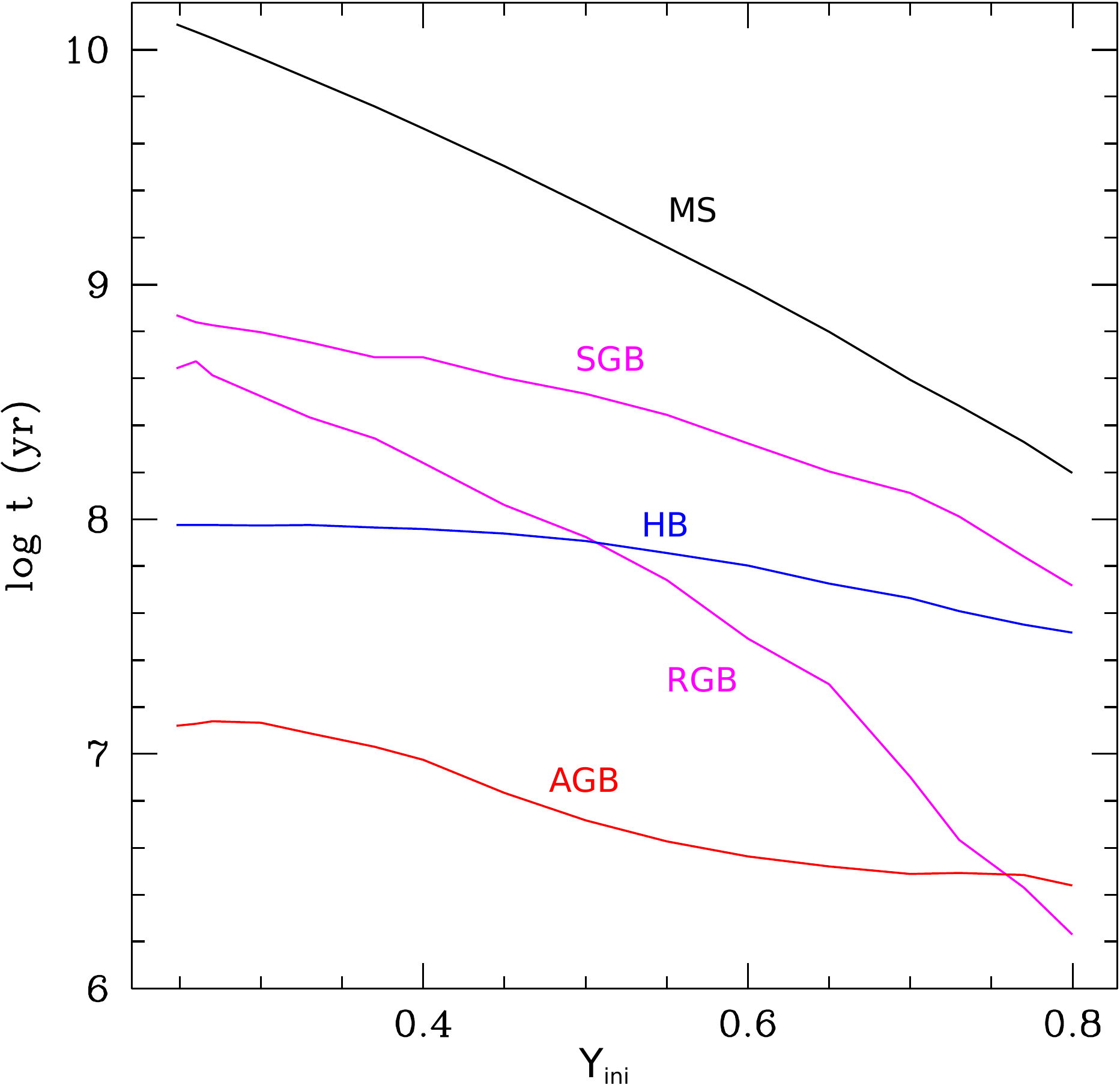}
\caption{Predictions for 0.8~M$_{\odot}$, [Fe/H] = -1.75 stellar models computed with different values of the initial He mass fraction Y$_{ini}$.
(Left) Evolution tracks in the HRD for three values of Y$_{ini}$. (Right) Duration of the different evolution phases as a function of Y$_{ini}$.
Figures from \citet{Chantereau15} }
\label{Fig:0p8_Yini_HRD}
\end{figure}

\begin{figure}
\centering
\includegraphics[width=0.4\textwidth]{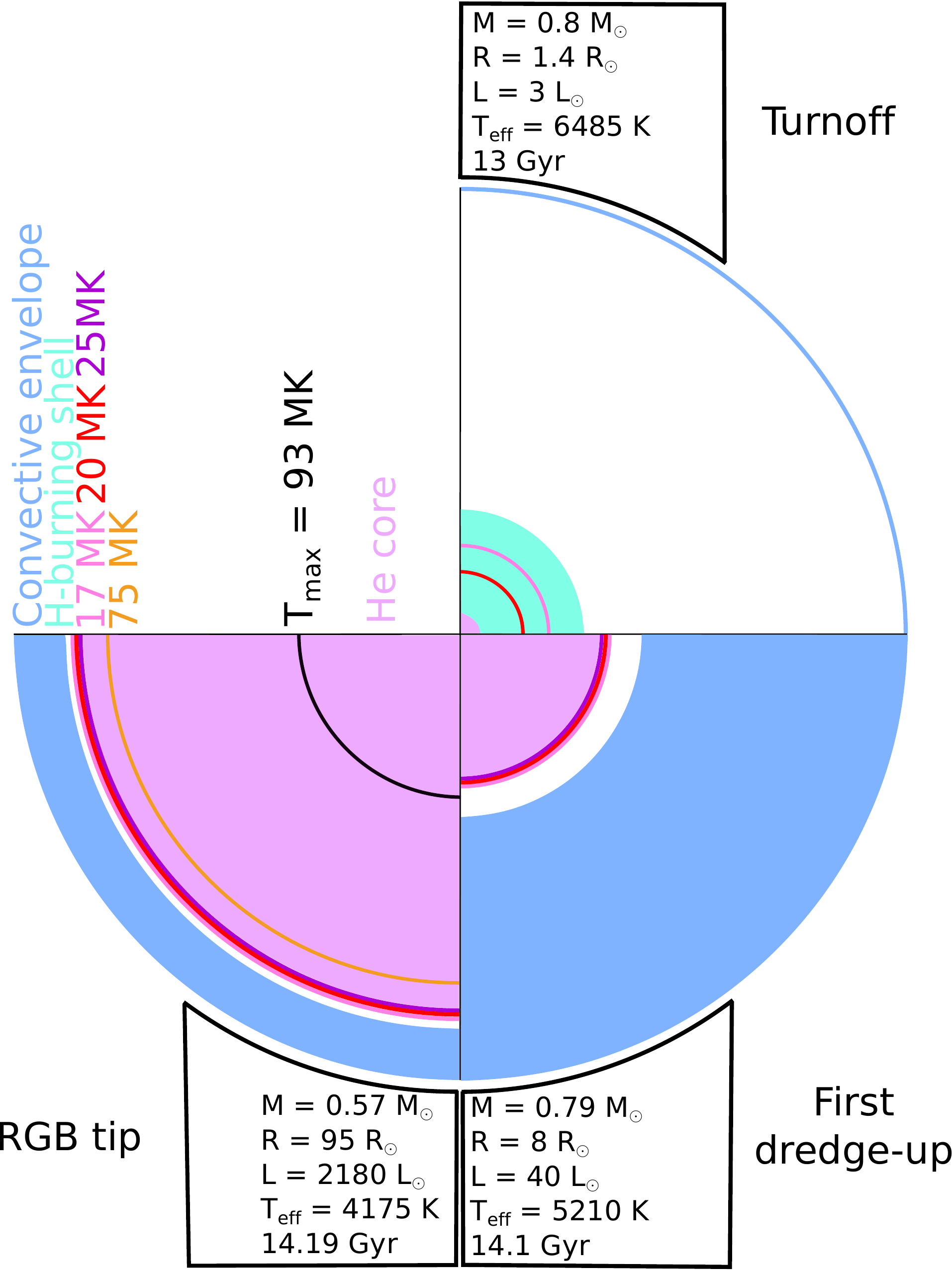}
\includegraphics[width=0.4\textwidth]{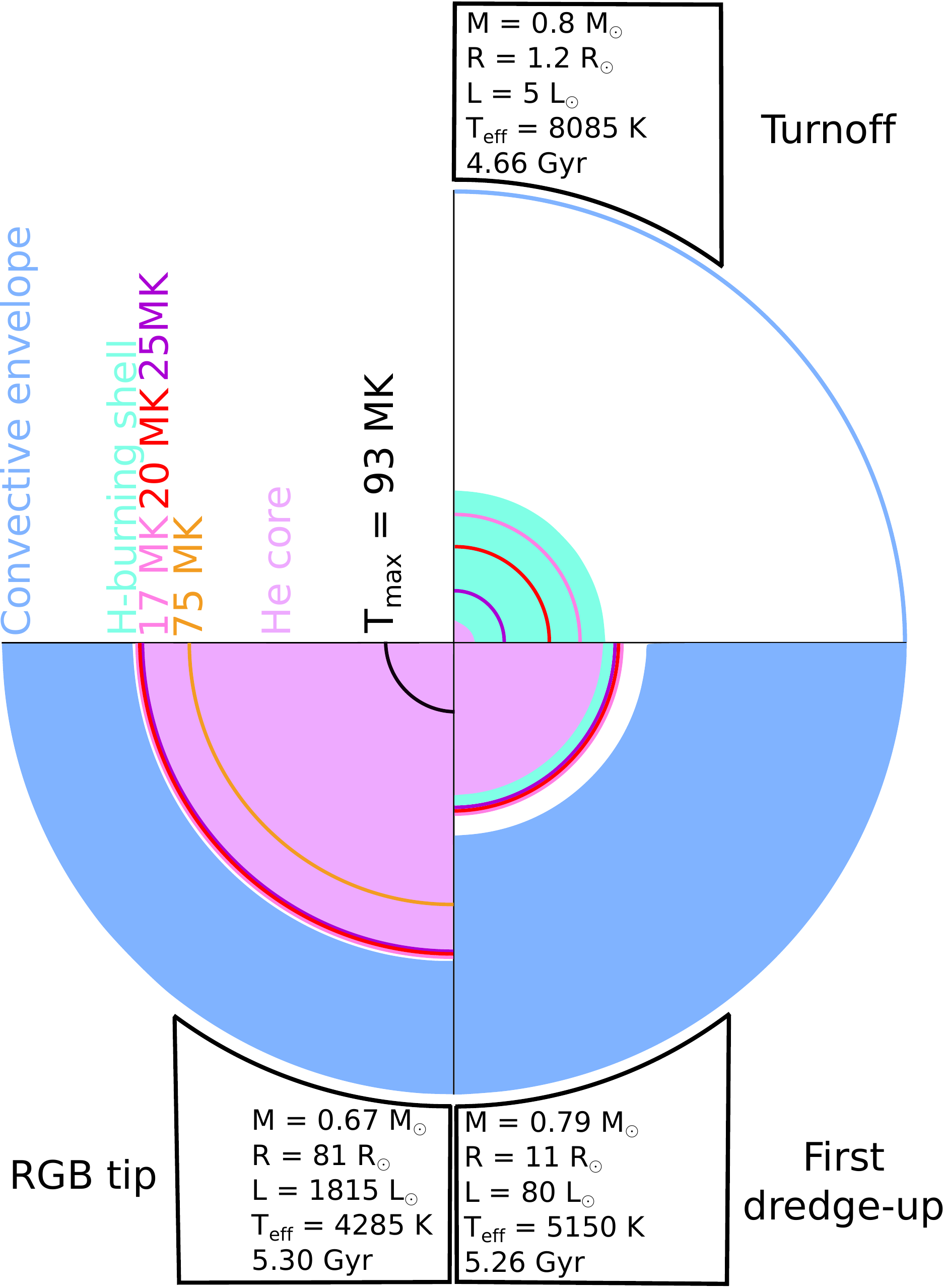}
\caption{Global characteristics of the 0.8~M$_{\odot}$, [Fe/H] = -1.75 stellar models computed with Y$_{ini}$ = 0.248 and 0.4 (left and right respectively), at peculiar stages of the evolution: Main sequence turnoff, end of the first dredge-up (at the time on the RGB where the convective envelope is the deepest), and tip of the RGB (at the time of the core He-flash). At each phase we give the total stellar mass and radius, its luminosity, effective temperature, and age. The size of the convective envelope (blue), of the H-burning shell (green), and of the He-core (magenta) are plotted as a function of the current total stellar mass. The other colors indicate the location of different isotherms (17, 20, 25, 75 million Kelvin; maximum temperature T$_{max}$ in black), also as a function of the current total stellar mass.
Figures from Chantereau (2016, PhD thesis, Geneva)}
\label{Fig:0p8_Yini_structure}
\end{figure}

Numerous papers have been devoted to the study of the impact of the initial helium content of a star on its evolution \citep{Iben68,HartwickvdB73,Lee78,SweigartGross78,D'Antona02,D'Antona05,D'AntonaCaloi04,D'AntonaCaloi08,Salaris06,Bertellietal08,Pietrinferni09,D'Antona10,Valcarce12,Karakas14,Dotteretal15,Chantereau15,Chantereau16}.
We describe and illustrate the main points below, using predictions of the theoretical models of \citet[][]{Chantereau15} computed with the stellar evolution code STAREVOL for [Fe/H] = -1.75 and the initial He mass fraction Y$_{ini}$\footnote{X$_{ini}$+Y$_{ini}$+Z$_{ini}$=1, with X$_{ini}$, Y$_{ini}$, and Z$_{ini}$ the initial mass fractions in hydrogen, helium, and metals, respectively} varying between 0.248 and 0.8. Mass loss is accounted for according to the prescription of \citet[][with the $\eta$ parameter equal to 0.5]{Reimers75}  up to the end of central He-burning and  of \citet{Vassiliadis93} in the more advanced evolution phases. 
For more details on the input physics of the models, we refer to Chantereau (2016, PhD thesis, Geneva).  

We focus first on the case of a 0.8~M$_{\odot}$ star. We show in Fig.~\ref{Fig:0p8_Yini_HRD} its evolution paths in the Hertzsprung-Russell diagram for three values of Y$_{ini}$ (0.248, which is the canonical value for the considered [Fe/H], 0.4, and 0.8). 
We quantify in Fig.~\ref{Fig:0p8_Yini_structure} some key properties of the corresponding models for Y$_{ini}$ = 0.248 and 0.4 at three different evolution stages (main sequence turnoff, maximum of the first dredge-up on the RGB, and RGB tip).
Basically, a larger initial He content modifies two important properties of the stellar gas: It reduces the opacity (Thomson scattering), and it increases the mean molecular weight of the material. Therefore, a star of a given initial mass is more luminous, hotter, and more compact when Y$_{ini}$ is higher. During the main sequence, the central temperature increases and the internal temperature profile steepens for increasing Y$_{ini}$. This impacts the mode of central H-burning: Nuclear burning proceeds mainly through the pp-chains in a radiative core when Y$_{ini}$ is lower than $\sim$ 0.5, and through the CNO-cycle in a convection core for higher Y$_{ini}$ values.
Consequently, the main sequence lifetime for the 0.8~M$_{\odot}$ models decreases from 12.9~Gyr to 4.66~Gyr and 158~Myr when Y$_{ini}$ increases from 0.248 to 0.4 and 0.8 respectively (Fig.~\ref{Fig:0p8_Yini_HRD}) . 

The following evolution phases are also affected. In particular, stars with higher Y$_{ini}$ spend less time on the red giant branch, 
and the mass of the He-core as well as the total stellar luminosity at the RGB tip decrease as a result of lower degeneracy and higher temperature. 
Interestingly, 0.8~M$_{\odot}$,  [Fe/H] = -1.75 stars with Y$_{ini}$ higher than $\sim$ 0.5 ignite central He-burning smoothly and at relatively low luminosity, while those with lower   Y$_{ini}$ undergo the so-called helium flash at very high luminosity at the tip of the red giant branch. 

Consequently, the duration of the transition between central H- and He-burning is strongly reduced when  Y$_{ini}$ increases (1.18~Gyr, 663~Myr, and 53.8~Myr after the main sequence turnoff for Y$_{ini}$ equal to 0.248, 0.4, and 0.8 respectively). This  modifies the amount of mass the star loses through winds along the RGB (it decreases with increasing Y$_{ini}$). This affects directly the total mass of the star as well as the ratio between the mass of the He-core and that of the stellar envelope, thus the effective temperature of the star when it arrives on the horizontal branch. The stars that ignite He through flashes at the RGB tip (Y$_{ini}$ lower than $\sim$ 0.5) all arrive on the HB with very similar He-core mass, thus the duration of the HB is not strongly affected, at odds with more He-enriched stars that stay on the HB for a very short time (HB duration is 94, 91, and 33~Myr for Y$_{ini}$ equal to 0.248, 0.4, and 0.8 respectively).

Finally, all the 0.8~M$_{\odot}$,  [Fe/H] = -1.75 models computed with Y$_{ini}$ lower than $\sim$ 0.6 - 0.65 reach the thermally-pulsing phase on the asymptotic giant branch (TP-AGB). The number of thermal pulses (TP) increases when Y$_{ini}$ varies from 0.248 to 0.45, because of the larger total stellar mass when the first TP occurs. Then the number of TP decreases when Y$_{ini}$ is higher than 0.45, because of the increasing mass ratio between the CO core and the convective envelope. When Y$_{ini}$ is higher than $\sim$ 0.6 - 0.65, the 0.8~M$_{\odot}$ models do not undergo any thermal pulse; they totally lose their envelope during their very short climbing of the early-AGB, and they evolve directly towards the white dwarf stage. Last but not least, the mass of the CO white dwarf remnant increases with increasing Y$_{ini}$ (0.57, 0.66, and 0.75~M$_{\odot}$ respectively for the 0.8~M$_{\odot}$,  [Fe/H] = -1.75 models computed with Y$_{ini}$ equal to 0.248, 0.4, and 0.8). 

\subsection{Impact of the initial helium content as a function of the initial stellar mass}
\label{subsection:HeenrichmentvsMini}

\begin{figure}
\centering
\includegraphics[width=0.6\textwidth]{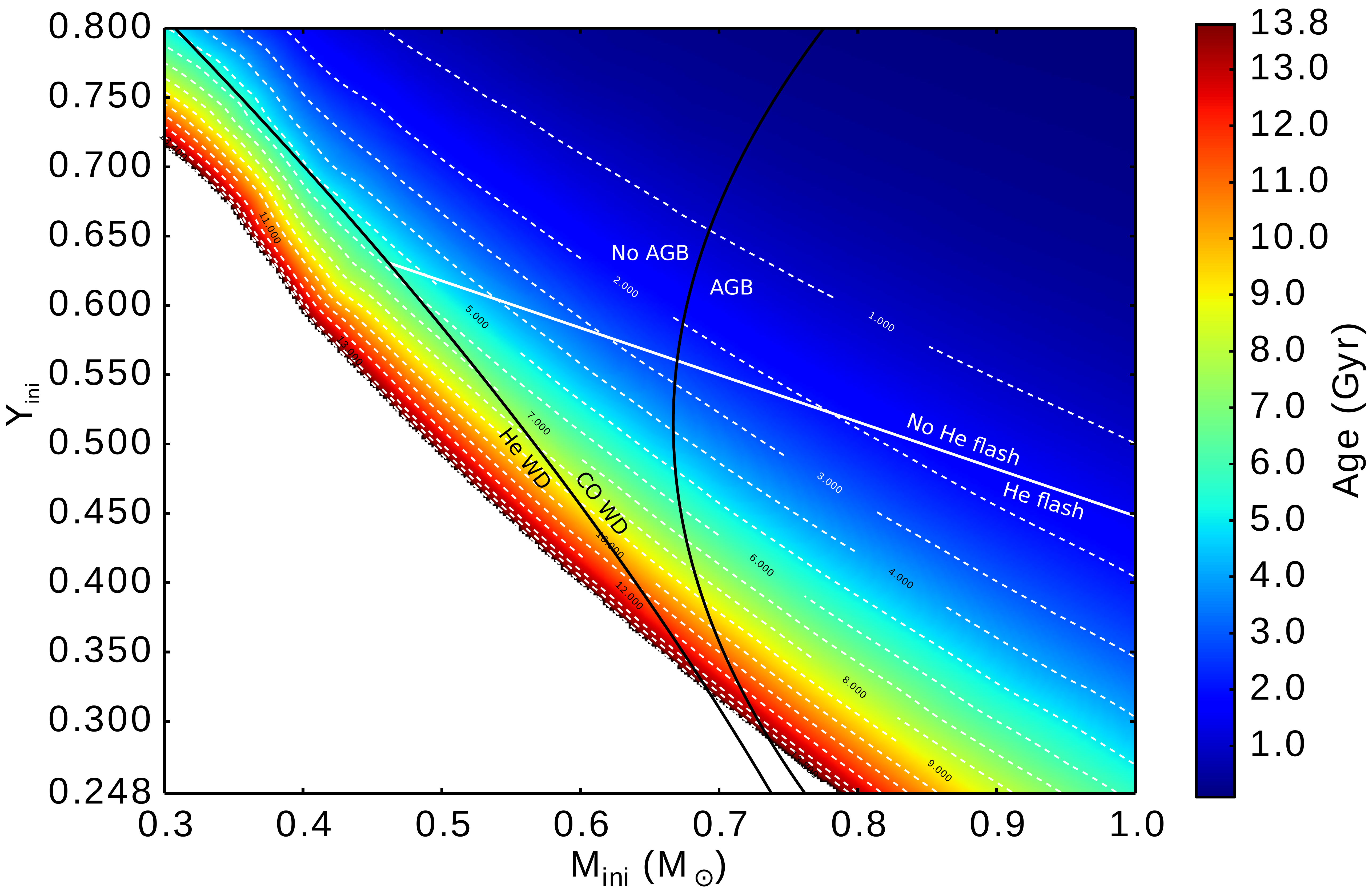}
\caption{This figure summarizes several effects induced by helium enrichment (Y$_{ini}$, ordinate) on the evolution of a star as a function of its mass (M$_{ini}$, abscissa). The predictions correspond to models computed with [Fe/H] = -1.75. Colors vary as a function of the age of the models at the main sequence turnoff (white dashed lines are isochrones with ages in Gyr, stoping at 13.8~Gyr). The white line delimits the (M$_{ini}$,Y$_{ini}$) area where He ignition (when it occurs) stars in non-degenerate conditions or with the so-called He-flash at the tip of the RGB. The black lines delimit the domains where stars become He or CO white dwarfs and where the stars climb or not the TP-AGB. 
Figure from \citet{Chantereau15} }
\label{Fig:allmasses_Yini_vaying}
\end{figure}

Figure~\ref{Fig:allmasses_Yini_vaying} summarizes all the effects of initial He variations described above for stars of different initial masses (models at [Fe/H] = -1.75 from \citealt{Chantereau15}). In addition to the impact on the stellar lifetimes, we emphasize the (M$_{ini}$,Y$_{ini}$) areas where the stars start central He-burning with a flash or not, climb the TP-AGB or become AGB-manqu\'e, and become He or CO white dwarfs.
No He-WD are expected to be present in GCs as a result of normal evolution path for canonical values of He (Y$_{ini}$ = 0.248), because the lifetime of their low-mass progenitors are significantly longer than the Hubble time (He-WD might however form in close binary systems due to extreme mass loss events; \citealt{Webbink75,Strickler09}). However, an He-WD population could be present in GCs as a result of the faster evolution of stars with Y$_{ini}$ higher than $\sim$ 0.33. Additionally, the initial mass for a star to become a white dwarf decreases with increasing helium.

\subsection{Consequences for synthetic globular clusters}
\label{syntheticCGs}

The initial helium content of a star strongly influences its properties, evolution path, lifetime, and fate, as shown previously. 
This must be accounted for to understand the characteristics of the multiple stellar populations in GCs, and to interpret the observed color-magnitude diagrams properly. 
This is also particularly important to pinpoint the nature of the polluters that have contributed to GC self-enrichment. 

Here we summarize the results of \citet{Chantereau16} who studied the characteristics at different ages of a stellar population born with Y$_{ini}$ varying between 0.248 and 0.8 (using the models of \citealt{Chantereau15} for [Fe/H] = -1.75 presented above). They assume the relation between the initial helium and sodium mass fraction predicted by the FRMS scenario  (\S~\ref{subsection:FRMSnucleosythesis}) that reproduces the observed distribution of Na in the GC NGC~6752. To construct their population synthesis models, \citet{Chantereau16} assume for the IMF the present day stellar mass function derived by \citet{Paresce00} and \citet{Salpeter55} IMF for stars initially less and more massive than 0.85~M$_{\odot}$ respectively. They neglect the dynamical effects that can lead to the ejection of stars from the cluster as well as the effects of stellar multiplicity (e.g. mass transfer in close binaries). They follow the evolution of an initial population of 300'000 stars with 21$\%$ having Y$_{ini}$ higher than 0.4. 

At the ages of 9 and 13~Gyr respectively (typical age range of Galactic GCs), only 75 and 67$\%$ of the initial sample stars are still alive (i.e., did not become dark remnant yet),  as a result of stellar evolution. The proportion of He-rich stars decreases with time, as these objects evolve faster. At 9 and 13~Gyr, the percentage of stars born with Y$_{ini}$ higher than 0.4 are 12 and 10$\%$ respectively (compare to the initial 21$\%$).
In the same model at 13.4 Gyr, 95\% of the low-mass stars lying two magnitudes below the turn-off were born with Y$_{\rm ini}$ between 0.248 and 0.4, and only 5 \% with Y$_{\rm ini}$ higher than 0.4. 
The fraction of He-rich stars decreases in the more advanced evolution phases. In particular, no star born with Y$_{\rm ini}$ higher than 0.4 should be found today on the horizontal branch, in agreement with current interpretations of GC horizontal branch morphologies in connection with the so-called second parameter problem (e.g. \citealt{D'AntonaCaloi04,Dotter10}). 

Finally, the correlation between the initial He and Na content given by the FRMS scenario and used by \citet{Chantereau16} 
provides a straightforward explanation to the lack of sodium-rich 2P AGB stars in GCs like NGC~6752 (\citealt{Charbonnel13}; observations by \citealt{Campbell13}) due to the impact of the initial helium content on the stellar lifetime and evolution path in the Hertzsprung-Russel diagram. They also account for the finding of different sodium spreads that are observed along the AGB in the Galactic globular clusters of different ages and [Fe/H] values \citep{CharbonnelChantereau16,Wangetal16}. 

\subsection{From the theoretical to the observational planes}
\label{syntheticCGs}

The maximum helium spread derived to date for a GC is 0.13~dex (in mass fraction, i.e., Y$_{\rm ini}$ between 0.248 and 0.378), and it was found in NGC~2808 (\citealt{Miloneetal12}; \S~\ref{obshelium}). Other GCs seem to show less He spread, as derived with the isochrone fitting method. In NGC~6752, a maximum He spread of 0.03~dex (in mass fraction) was derived by \citet{Milone13}. 
This clearly poses a problem to the polluters proposed in the literature \citep[e.g.][]{Bastianetal15}, which all fail to account for such extremely low helium enrichment (but see 
\S~\ref{subsection:Supermassivestarsnucleosythesis} for the case of supermassive stars).

As of today however, a direct and reliable comparison between theoretical predictions and observed GC CMDs is still difficult, for several reasons. 
Going from the theoretical to the observational plane requires indeed the use of model atmospheres and temperature-color transformations suited for the proper helium range and associated peculiar composition of 2P stars. This has already been done up to  a certain extent \citep{Sbordone11,Milone13,Cassisi13}, but not for large helium abundance spread, as such tools are not available yet.
In addition, stellar models for 1P and 2P long-lived low-mass stars should include the effects of atomic diffusion, rotation, and various hydrodynamical processes, that are known to have a non negligible impact on stellar evolution and lifetimes, and on the position of stars within the HRD. 
Clearly, more fundamental work needs to be done before we can definitively understand MSP formation and evolution in GCs.

\section{Conclusions}
\label{section:multiplepopulationsopenquestions}

Extensive literature is devoted to the ubiquitous presence of MSPs in GCs. 
As of today, none of the current scenario for GC self-enrichment is able to explain the spectroscopic and photometric complexity and variety of these systems.
Crucial questions remain open. 

What kind of short-lived massive stars did pollute the intracluster material with only hydrogen-burning products during the early GC infancy? The most recent studies point to massive and supermassive stars. The evolution of these objects and their interactions with the intracluster gas and dust in very compact environments appears to be a crucial ingredient of this very complex problem \citep{Krauseetal12,Krauseetal13,Krauseetal16}, but many aspects remain to be explored. 
What has been the chemical and dynamical impact of massive and supermassive stars on their host clusters? How did the feedback with the intra-cluster medium impact the formation and evolution of stars on various temporal and spatial scales, as well as the dynamics of the entire clusters?
Did they contribute to gas ejection eventually associated to a significant loss of stars without disrupting the cluster by a drastic change of potential well? 

How did the multiple populations of low-mass stars that are still alive today form in each individual GC, how did they inherite their chemical peculiarities? What are the consequences for their evolution and fate? 
Why are these MSPs not observed in other types of star clusters, like open clusters? 
Does this reflect very specific star formation mechanisms in the early universe? 
Or will we also find similar MSPs in young massive star clusters formed in present-day conditions once we will have the suited observational tools (Bastian, this volume)?
What was the original mass of GCs, and their contribution to galactic haloes (Baumgardt, this volume)?

In view of the complexity of the problems raised by the presence of multiple stellar populations, GCs will remain for long extraordinary exploration territories for stellar physics, evolution, and dynamics. Finding answers will require exchange of ideas and close collaboration of astrophysicists with observational, theoretical, and numerical expertise in stellar evolution, interstellar matter magnetohydrodynamics, stellar dynamics, formation and evolution of galaxies, cosmology, multidimensional numerical simulations, N-body simulations, and multi-wavelength high-precision photometry, spectroscopy, and astrometry. Thanks to their collective efforts, a robust and consensual explanation will hopefully be found before the organisation of the next EES on the topic. 
 
\section*{Acknowledgements}
I dedicate this paper to my dear friend, inspiring mentor, and respected colleague, Jean-Paul Zahn, who has been the director of the Evry Schatzman School of PNPS/CNRS/INSU from 1997 to 2009. I warmly thank my closest collaborators with whom I worked on several problems related to GCs, Martin Krause, Nikos Prantzos, William Chantereau, Thibaut Decressin, Georges Meynet, Francesca Primas, Holger Baumgardt, Daniel Schaerer, and more recently Nate Bastian, Mark Gieles, Vincent H\'enault-Brunet, Fabrice Martins, and Yue Wang. I am grateful to Franca D'Antona, Paolo Ventura, Pavel Denissenkov, and Andr\'e Maeder for lively discussions on nucleosynthesis in stars.
I acknowledge support from the Swiss National Science Foundation (FNS) for the project 200020-159543 ``Multiple stellar populations in massive star clusters: Formation, evolution, dynamics, impact on galactic evolution" (PI C.C.). I thank the International Space Science Institute (ISSI, Bern, CH) for welcoming the activities of ISSI Team 271 ``Massive star clusters across the Hubble Time (2013 - 2016, team leader C.C.).

\bibliographystyle{astron}
\bibliography{Bibliography}

\end{document}